\documentclass[twocolumn,showpacs,amsmath,amssymb,superscriptaddress]{revtex4}
\usepackage{amsmath,amssymb,color,latexsym,natbib,graphicx}
\usepackage{ulem,color,soul}
\usepackage{dcolumn}
\usepackage{bm}
\usepackage{subfigure}
\usepackage{graphicx}
\usepackage{caption}
\newcommand{\q}[1]{``#1''}
\usepackage{modroman}
\usepackage[sort&compress]{natbib}

\begin{document}

\preprint{1}

\title{Coexistence of weak and strong wave turbulence in incompressible Hall magnetohydrodynamics}
\author{Romain Meyrand}
\email{romain.meyrand@lpp.polytechnique.fr}
\affiliation{Space Sciences Laboratory, University of California, Berkeley, CA 94720, USA}
\affiliation{LPP, \'Ecole polytechnique, F-91128 Palaiseau Cedex, France}
\author{Khurom H. Kiyani}
\affiliation{LPP, \'Ecole polytechnique, F-91128 Palaiseau Cedex, France}
\affiliation{Centre for Fusion, Space and Astrophysics, University of Warwick, Coventry CV4 7AL, UK}
\author{\"Ozgur D. G\"urcan}
\affiliation{LPP, \'Ecole polytechnique, F-91128 Palaiseau Cedex, France}
\author{S\'ebastien Galtier}
\affiliation{LPP, \'Ecole polytechnique, F-91128 Palaiseau Cedex, France}
\affiliation{Universit\'e Paris-Sud, France}

\date{\today}
\begin{abstract}
We report a numerical investigation of three dimensional, incompressible, Hall magnetohydrodynamic turbulence with a relatively strong mean magnetic field. Using 
helicity decomposition and cross-bicoherence analysis, we observe that the resonant three--wave coupling is substantial among ion cyclotron and whistler waves. 
A detailed study of the degree of non-linearity of these two populations shows that the ion cyclotron component experiences a transition from weak to strong wave 
turbulence going from large to small scales, while the whistler fluctuations display a weak wave turbulence character for all scales. This non-trivial coexistence of the two 
regimes with the two populations of waves gives rise to anomalous anisotropy and scaling properties. The weak and strong wave turbulence components can be 
distinguished rather efficiently using spatio-temporal Fourier transforms. The analysis shows that while resonant triadic interactions survive the highly non-linear bath of 
ion cyclotron fluctuations at large scales for which the degree of non-linearity is low for both populations of waves, whistler waves tend to be killed by the non-linear cross-coupling at smaller scales where the ion cyclotron component is in the strong wave turbulent regime. Such situation may have far-reaching implications for the physics of magnetized turbulence in many astrophysical and space plasmas where different waves coexist and compete to transfer non-linearly energy 
across scales.
\end{abstract}
\maketitle
\section{Introduction} 
A sea of weakly non-linear dispersive waves can exchange energy among spatial scales resulting in a highly turbulent state. Because of the weak 
non-linear coupling, the energy transfer takes place mainly {\it via} resonant interactions among a set of waves. The resulting behaviour for a statistically stationary 
state, far from thermodynamic equilibrium, bears resemblance to the cascade picture in three-dimensional (3D) hydrodynamic turbulence: energy 
injected at large scale cascades toward smaller scales where dissipation tranforms it irreversibly into heat. A fundamental difference with strong hydrodynamic 
turbulence is that an out-of-equilibrium system made of weakly interacting waves is free of the closure difficulties and thus appears to be a solvable 
problem for which rigorous analytical predictions can be made in the framework of the Weak Turbulence Theory (WTT) \cite{Benney67}. 
WTT was developed during the sixties with the pioneering works on gravity waves travelling at the surface of the ocean \cite{Hasselmann62}. Soon after, the 
theory was applied to plasma physics \cite{Vedenov67} and since then, to an increasing number of physical problems, ranging from quantum \cite{Nazarenko06} to 
astrophysical scales \cite{Micha03}.  

While the idealization described by WTT deals only with weakly non-linear random waves, in most of real situations  
strongly non-linear coherent structures coexist with incoherent weakly non-linear waves. In 3D magnetohydrodynamic (MHD) turbulence, because 2D vortices play a 
catalytic role for the triadic interactions of Alfv\'en waves, non-linear coherent structures have a strong impact on the weak wave turbulent dynamics 
\cite{Meyrand15}.  A similar situation is also encountered in non-equilibrium Bose-Einstein condensation \cite{Lvov03} or in optical turbulence \cite{Dyachenko92}. 
In these cases, however, it appears as a consequence of the inverse cascade of energy. Another possible complication is that the weak wave turbulent cascade 
can bring energy all the way 
down to the scales where the dominant forces may change the nature of the waves. This is the case, for example, at the surface of water where gravity waves 
transform at small scale into capillary waves \cite{Aubourg15} under the influence of surface tension. Similarly in rotating turbulence, the effect of the Coriolis force 
which decreases as a function of scale, may lead to the conversion of inertial waves into highly non-linear fluctuations. 
Because of these \q{real life} effects, experiments often show deviations from the existing predictions and weak wave turbulence is rarely observed in its pure form 
\cite{Mordant08}. However, in these examples the coherent non-linear structures and/or the different weak wave fields do not \q{live} in the same area of the spectral 
space which facilitates the analytical and experimental disentangling of the two components.

In plasmas physics, the situation seems to be different because an entire zoo of different waves can effectively coexist at any given scale. 
The question, then is: how does these different waves coexist and compete to transfer the energy across scales?
In this paper, we address this question using a simple archetypal example of plasma turbulence modelled by the incompressible Hall 
MHD equations. In the framework of Hall MHD, the electrons are assumed to be inertia-less and the electric field is determined by the equation of motion of the electron 
fluid. As a result the magnetic field is tied to the electrons  (modulo Ohmic losses) and not to the bulk fluid.  At sub-ion scales, this decoupling leads to the emergence of 
two circularly polarized waves with opposite polarity, the so-called whistler and ion cyclotron waves. 
The kinetic equations for three-wave interaction processes describing the non-linear dynamics of weak wave Hall MHD turbulence were derived in the general 
case, ie. including the non-linear interactions between different types of waves. The exact power law solutions were also derived analytically \cite{Galtier06} but only in 
the simplified case where the interactions between the ion cyclotron and whistler waves are negligible.
The aim of the article is to study in detail, to what extent this assumption is justified.

In section \ref{sec1} we describe the incompressible Hall MHD approximation and present the associated equations. We discuss the relevance of this appealing 
model in the context of magnetized plasma turbulence but also its limitations. \\
Section \ref{sec2} contains a tutorial on weak wave incompressible Hall MHD turbulence, a necessary background to what will follow. To avoid obscuring the physics with 
the complexity of the WWT formalism we provide an heuristic description able to recover the essential physics underlying the rigorous 
analytical prediction. In this section we introduce the generalized Els\"asser variables and the complex helical decomposition which will be used extensively thereafter.\\
In section \ref{sec3} we derive new analytical relationships at the level of the kinematics which allow to understand some numerical results which are presented 
subsequently. We show in particular that the velocity and magnetic field fluctuations spectra of same polarity are automatically linked. This section is rather technical and 
an impatient reader can skim through it without losing the essence of the article. \\
Section \ref{sec4} constitutes the core of the article. We present results from high resolution 3D direct numerical simulations.  We show especially that the ion cyclotron waves experience a transition from weak to strong wave turbulence going from large to small scales, while the whistler fluctuations display a weak wave turbulence character for all scales. This generates anomalous anisotropy and scaling properties. Those results raise fundamental questions about the applicability of WTT in the context of Hall MHD. Note that in the present study, we consider only sub-ion scales. Therefore large scales do not refer to MHD scales in this article. \\
Section \ref{sec5}  is dedicated to this latter issue. Using higher-order polyspectra techniques we show two fundamental results. First, resonant triadic interactions 
survive the highly non-linear bath of ion cyclotron fluctuations. Second, three--wave coupling is substantial among ion cyclotron and whistler waves. \\
In section \ref{sec6} we focus on the properties of the space-time Fourier spectrum. This study allows us to show that if resonant triadic interactions are at work at 
large scales for which both whistler and ion cyclotron fluctuations are weakly non-linear, whistlers are killed by local cross-coupling with strongly non-linear ion cyclotron 
fluctuations at smaller scales.\\
Section \ref{sec6}  is devoted to summarising this and other findings and to discussing their implications.   

\section{Incompressible Hall MHD} \label{sec1}
Hall MHD is a theoretical paradigm which captures both the MHD behaviour at long wavelengths and some of the kinetic effects that become important at small scales due 
to the decoupling between the electron and ion flows. This can be done by keeping the Hall current term in the ideal Ohm's law (in SI unit): 
\begin{equation}
\textbf{E}+\textbf{u}\times \textbf{B} - \dfrac{\textbf{j} \times \textbf{B}}{ne} - \dfrac{\nabla p_{e}}{ne} = 0 ,
\label{Ohm}
\end{equation}
 where $\textbf{E}$ is the electric field, $\textbf{u}$ the bulk velocity, $\textbf{j}$ the electric current, $\textbf{B}$ the magnetic field, $n$ the electron density, $e$ the 
 magnitude of the electron charge and $p_{e}$ the electron pressure.  The Hall term becomes dominant at length scales smaller than the ion inertial length $d_{i}$ 
 ($d_{i}\equiv c/\omega_{pi}$ with c being the speed of light and $\omega_{pi}$ being the ion plasma frequency) and time scales of the order of, or shorter than, 
 the ion cyclotron period $\omega_{ci}^{-1}$. 
 
The linear dispersion relation of the Hall MHD can be recovered exactly from the full kinetic dispersion relation in the limit of $T_{i} \ll T_{e}$, $v_{thi}\ll \omega/ |k_{\parallel}|\ll 
v_{the}$ and $\omega\ll\omega_{ci}$ \cite{Ito04, Hirose04}, with $T_{i,e}$ and $v_{thi, the}$ respectively the ion/electron temperature and thermal speed. The cold ions 
assumption permits a finite ion inertial length with a vanishing ion Larmor radius $\rho_{i}$ ($\rho_{i}\equiv v_{thi}/\omega_{pi}$). Thus, Hall MHD keeps finite 
frequency effects, while neglecting finite Larmor radius effects. The second approximation ensures that the ions and the electrons are respectively cold and hot enough to not be subjected to Landau damping, while the third approximation corresponds to neglect the ion cyclotron resonance. If the approximation $k_{\parallel}\ll k_{\perp}$ is further 
considered then Hall MHD can be derived from the cold ion limit of gyrokinetics \cite{Schekochihin09}. Note, however, that for $k_{\parallel}$ not small compared to 
$k_{\perp}$, the gyrokinetics is not valid, while Hall MHD continues to describe the cold-ion limit correctly capturing in particular the whistler branch of the dispersion 
relation \cite{Sahraoui07,Schekochihin09}. 
In other words, Hall MHD is a valid approximation at the condition that the parameter regime considered is such that it is rigorously justified to completely ignore 
collisionless damping and finite Larmor radius effect. If taken literally, these restrictive ordering excludes \textit{de facto} many plasmas of interest. Examples of plasmas 
where it may hold are cold and dense regions of protoplanetary discs \cite{Kunz13}, crusts of neutron stars \cite{Gourgouliatos14} or some plasma research devices like the The 
Madison Plasma Dynamo experiment \cite{Cooper14} or the Wisconsin Plasma Astrophysics Laboratory \cite{Forest15}.  However, one may legitimately expect that the 
plasma behaviour captured by Hall MHD will qualitatively hold beyond its rigorous limits of applicability. MHD is a notorious example of a description known to work 
rather well far outside its strict limits of validity as is it often the case with many other simplified plasma models \cite{biskamp2003, Loureiro15}.

In the context of solar wind turbulence the cold ions limit has been fingered as an important limitation of Hall MHD because spurious undamped wave modes appear when 
the ions temperature is finite \cite{Howes06} and because the spectral break and the associated change in the nature of the turbulent cascade in Hall MHD turbulence 
may appear at the wrong scale ($d_{i}$ instead of $\rho_{i}$) \cite{Schekochihin09}. 
Concerning the first argument, it is important to recall two points. First, robust Kolmogorov-like power-law spectra of compressible fluctuations in the inertial range of 
solar wind turbulence are frequently observed \cite{Marsch90, Kellogg05} even though Landau damping of such fluctuations should be noticeable at these scales 
\cite{Schekochihin09}. The same phenomenon has been observed at sub-ion scales \cite{Sahraoui09, Alexandrova09} and in some kinetic simulations \cite{Chang11}. 
These observations show that collisionless damping rates derived from linear kinetic theory are not applicable in a turbulent plasma. Second, because of the anisotropy, 
the (linearly damped) compressive perturbations have a tendency to be passively mixed by the undamped Alfv\'enic turbulence \cite{Maron01}. Consequently even if the 
linear kinetic theory was applicable to plasma turbulence, the existence of certain (undamped) wave modes in Hall MHD that are linearly damped in a weakly collisional 
plasma would not necessarily affect the non-linear dynamics. 
Concerning the second argument, \textit{in situ} measurements of magnetic fluctuations in the solar wind show that the spectral break at ion kinetic scales occurs either 
at $d_{i}$ or $\rho_{i}$ depending of the value of $\beta_{i}$ \cite{Chen14} showing that different physical processes are at work and that it is difficult to lock up this 
complex phenomenon in a complete and exclusive theory. More generally, if one considers that the salient feature of plasma turbulence is a flux of invariants through scales 
rather than thermodynamic potentials like temperature, it might appear that a fluid model like Hall MHD can provide useful insights in the study of plasma turbulence 
without bringing in the full complexity of the kinetic theory.  

The (inviscid and ideal) incompressible 3D Hall MHD equations can be obtained from incompressible MHD if one introduces the  generalized Ohm's law 
(\ref{Ohm}) into Maxwell-Faraday's equation and by assuming that the electron pressure $p_{e}$ is a scalar (this can be justified in the collisional limit or in the isothermal 
electron fluid approximation \cite{Schekochihin09}). It gives: 
\begin{eqnarray}
\nabla\cdot\textbf{u} &=& 0 , \quad\nabla\cdot\textbf{b} = 0 , \label{hmhd1}\\
\dfrac{\partial\textbf{u}}{\partial t} + \textbf{u}\cdot\nabla \textbf{u} &=& -\nabla P_{*} + \textbf{b}_{0}\cdot\nabla \textbf{b}+\textbf{b}\cdot\nabla \textbf{b}  ,
\label{hmhd2} \\
\dfrac{\partial\textbf{b}}{\partial t} + \textbf{u}\cdot\nabla \textbf{b} &=& (\textbf{b}_{0}\cdot\nabla)(\textbf{u} - 
d_{i}\nabla\times\textbf{b})+\textbf{b}\cdot\nabla\textbf{u}\nonumber\\
&& -d_{i}\nabla\times\left[(\nabla\times \textbf{b})\times \textbf{b} \right] , \label{hmhd3} 
\end{eqnarray}
where $P_*$ is the total pressure, $\textbf{b}$ is the magnetic field normalized to a velocity ($\textbf{b} = \textbf{B} /\sqrt{\mu_0 n m_i}$, 
with $m_i$ the ion mass) and $\textbf{b}_{0}$ is a uniform normalized magnetic field. The assumption of incompressibility allows dropping 
of the sonic wave which is thought to be less relevant due to its damping by kinetic effects \cite{Hunana11}, while describing accurately the two remaining 
dispersive branches of compressible Hall MHD at finite $\beta$ values \cite{Galtier15}. Note that equation (\ref{hmhd3}) in the limit $kd_{i}\gg 1$ and $k_{\perp}\gg 
k_{\parallel}$ is mathematically similar to the electron reduced MHD (ERMHD) equations to within a constant coefficient probably not essential for qualitative models of 
turbulence \cite{Schekochihin09}. Thus, incompressible Hall MHD is useful for understanding kinetic Alfv\'en wave cascade too. 
This point is non-trivial and remarkable in that the latter is indeed fundamentally linked to the compressible nature of the ion flow. 

\section{Heuristic description of Hall MHD WTT}\label{sec2}
The study of weak wave turbulence in Hall MHD is a difficult task requiring great analytical efforts. Fortunately it is possible to derive a generalized heuristic description able 
to recover the essential physics underlying the rigorous analytical prediction of WTT \cite{Galtier06}. The first step is to introduce the generalized  Els\"asser 
variables adapted to Hall MHD. To do so, it turns out that it is necessary to use a complex helical decomposition. Such decomposition provides a compact description 
of the dynamics and allows diagonalization of the system dealing with circularly polarized waves. This approach was used to study the dynamics of helicity, inertial, whistler 
or magnetostrophic waves to cite only few examples \cite{Craya58, Moffat70, Kraichnan73, Waleffe92, Galtier03a, Galtier14}. 
The helicity decomposition is non other than a decomposition into right-handed and 
left-handed polarization states of plane waves. At the linear level it is an elegant tool to derive the wave properties like the dispersion relation. At the non-linear level, it 
provides a powerful method to tackle WTT problems, where the use of such a decomposition is necessary for the correct derivation of the asymptotic equations by the Eulerian methods \cite{Galtier06}. It is worth pointing out that the advantages of the helical decomposition transcend the 
analytical aspects, as it gives a more physically intuitive description of the problem enabling headway that would otherwise be probably too difficult to do. 
\subsection{Helicity basis}
The complex helicity decomposition is defined by:
\begin{equation}
\textbf{h}^{\Lambda}(\textbf{k})\equiv \textbf{h}_{\textbf{k}}^{\Lambda} = \hat{\textbf{e}}_{\theta} + i\Lambda\hat{\textbf{e}}_{\Phi},
\label{basis0}
\end{equation}
where $i^2 = -1$,
\begin{equation}
\hat{\textbf{e}}_{\theta} = \hat{\textbf{e}}_{\Phi} \times \hat{\textbf{e}}_{k}, \quad \hat{\textbf{e}}_{\Phi} = \dfrac{\hat{\textbf{e}}_{\parallel} \times \hat{\textbf{e}}_{k}}
{\vert \hat{\textbf{e}}_{\parallel} \times \hat{\textbf{e}}_{k}\vert}, 
\label{efi}
\end{equation}
and therefore $\vert\hat{\textbf{e}}_{\theta}\vert = \vert\hat{\textbf{e}}_{\Phi}\vert = 1$. In these relations, the wave vector $\textbf{k}= k\hat{\textbf{e}}_{\textbf{k}} = 
\textbf{k}_{\perp}+ \textbf{k}_{\parallel}\hat{\textbf{e}}_{\parallel}$  (with $k=\vert\textbf{k}\vert$, $k_{\perp}=\vert\textbf{k}_{\perp}\vert$, 
$\vert\hat{\textbf{e}}_{\textbf{k}}\vert=1$, $\hat{\textbf{e}}_{\parallel}$ being the direction along $\textbf{b}_{0}$). $\Lambda$ is called the wave polarization and takes 
the values $\pm$. We note in passing that $(\hat{\textbf{e}}_{\textbf{k}},\textbf{h}_{\textbf{k}}^{+},\textbf{h}_{\textbf{k}}^{-})$ forms a complex basis with the following 
properties:
\begin{eqnarray}
\textbf{h}_{\textbf{k}}^{-\Lambda} = \textbf{h}_{-\textbf{k}}^{\Lambda} ,\\
\hat{\textbf{e}}_{\textbf{k}}\times \textbf{h}_{\textbf{k}}^{\Lambda} = -i\Lambda \textbf{h}_{\textbf{k}}^{\Lambda} ,\\
\textbf{k}\cdot\textbf{h}_{\textbf{k}}^{\Lambda} = 0 \, ,\\
\textbf{h}_{\textbf{k}}^{\Lambda} \cdot \textbf{h}_{\textbf{k}}^{\Lambda^{'}} = 2\delta_{-\Lambda^{'}\Lambda}  .
\end{eqnarray}
With this decomposition we see that the incompressibility conditions (\ref{hmhd1}) are automatically satisfied. The Fourier transform of the original vectors $\textbf{u} 
(\textbf{x})$ and $\textbf{b} (\textbf{x})$ can be projected on the helicity basis; we write: 
\begin{eqnarray}
\hat{\textbf{u}}(\textbf{k}) =\sum_{\Lambda} \mathcal{U}_{\Lambda}(\textbf{k}) \textbf{h}_{k}^{\Lambda} \, , \label{ulambda}\\
\hat{\textbf{b}}(\textbf{k}) =\sum_{\Lambda} \mathcal{B}_{\Lambda}(\textbf{k}) \textbf{h}_{k}^{\Lambda} \, .\label{blambda}
\end{eqnarray}
\subsection{Eigenvectors and eigenmodes}\label{Eigenmodes}
The introduction of (\ref{ulambda})--(\ref{blambda}) into the Fourier transform of (\ref{hmhd2})--(\ref{hmhd3}) gives after the projection on ${\bf h^{\Lambda}_{-k}}$ 
and linearisation:
\begin{eqnarray}
\partial_{t}\mathcal{U}_{\Lambda}-ib_{0}k_{\parallel}\mathcal{B}_{\Lambda} &=& 0,\\
\partial_{t}\mathcal{B}_{\Lambda}-ib_{0}k_{\parallel}\mathcal{U}_{\Lambda} + i\Lambda d_{i}b_{0}k_{\parallel}k\mathcal{B}_{\Lambda} &=& 0.
\end{eqnarray}
To derive the dispersion relation, we  introduce the generalized Els\"asser fields (the eigenvectors):
\begin{equation}
\mathcal{Z}_{\Lambda}^{s} = \mathcal{U}_{\Lambda} + \xi_{\Lambda}^{s}\mathcal{B}_{\Lambda},
\label{zab}
\end{equation}
with $s=\pm$ and 
\begin{equation}
\xi_{\Lambda}^s (k) = \xi_{\Lambda}^s = - \dfrac{s d_{i}k}{2}\left(s\Lambda + \sqrt{1+\dfrac{4}{d_{i}^{2}k^{2}}} \right).
\end{equation}
Then, we obtain:
\begin{equation}
\partial_{t}\mathcal{Z}_{\Lambda}^{s} = -i\omega_{\Lambda}^{s}\mathcal{Z}_{\Lambda}^{s},
\label{zdispersion}
\end{equation}
with the dispersion relation:
\begin{equation}
\omega_{\Lambda}^{s} = -b_{0}k_{\parallel}\xi_{\Lambda}^{s}.
\end{equation}
Incompressible Hall MHD supports R and L circularly polarized waves which correspond to (oblique) whistler and ion-cyclotron waves respectively. We can easily check that 
in the small-scale limit ($k d_i \to + \infty$), we have $\xi_{\Lambda}^s \to -s \, d_i k$ for whistler waves ($\Lambda = s$), and $\xi_{\Lambda}^s \to -s/(d_i k)$ for 
ion-cyclotron waves ($\Lambda = -s$). In the large-scale limit ($k d_i \to 0$), we find $\xi_{\Lambda}^s \to - s$ and we recover the classical Els\"asser variables used in 
standard MHD. 
\subsection{Anisotropic Iroshnikov-Kraichnan spectrum of Hall MHD turbulence}\label{heuristic}
The anisotropic heuristic theory of Hall MHD WTT is given in \cite{Galtier06}. We will recall here the main steps of the derivation. 
The non-linear time built on the generalized Els\"asser variables can be written as:
\begin{equation}
\tau_{nl} \sim (k_{\perp}{\cal Z}_{\Lambda}^s)^{-1}.
\end{equation} 
Note here that ${\cal Z}_{\Lambda}^s$ has a dimension of a velocity, in other words, it is not taken in Fourier space as it was introduced in Section \ref{Eigenmodes}. 
The period of Hall MHD waves $\tau_{\omega}$ is given by 
\begin{equation}
\tau_{\omega}\sim (\omega_{\Lambda}^{s})^{-1}= -(b_{0}k_{\parallel}\xi_{\Lambda}^s)^{-1} .
\end{equation}
The characteristic transfer time of energy $\tau_{tr}$ can, as far as dimensional analysis is concerned, be an arbitrary function of these $8$ different times. Additional 
physical assumptions are therefore necessary to fix the scaling. In weak wave turbulence we have the inequality $\tau_{\omega}\ll \tau_{nl}$ and many stochastic 
collisions are necessary to modify significantly a wave packet. If we assume that the cumulative perturbation evolves as a random walk, the transfer time becomes 
\cite{Iroshnikov64, Kraichnan65} 
 \begin{equation}
\tau_{tr} \sim \tau_{nl}^{2}/\tau_{w}. 
 \end{equation}
If we now assume a stationary state for which the mean rate of energy dissipation per unit mass $\epsilon$ is independant of the scale we obtain
 \begin{equation}
 \epsilon \sim \dfrac{E}{\tau_{tr}}\sim \dfrac{E(k_{\perp}, k_{\parallel})k_{\perp}k_{\parallel}}{\tau_{tr}}\sim \dfrac{E(k_{\perp}, k_{\parallel})k_{\perp}^{3}{{\cal 
Z}_{\Lambda}^s}^{2}}{-B_{0}\xi_{\Lambda}^s},
  \end{equation}
which gives after some algebra
\begin{equation}
  E(k_{\perp},k_{\parallel})\sim \sqrt{\epsilon B_{0}}k_{\perp}^{-2}k_{\parallel}^{-1/2}(1+k_{\perp}^{2}d_{i}^{2})^{-1/4}.
  \label{heuristic_spectrum}
\end{equation}
 We recover, in the small scale limit ($k_{\perp}d_{i}\gg 1$), the expected scaling law for whistler as well as ion cyclotron wave turbulence  $E(k_{\perp},k_{\parallel}) \sim  
 k_{\perp}^{-5/2}k_{\parallel}^{-1/2}$ and, in the large scale limit ($k_{\perp}d_{i}\ll 1$), the Alfv\'en wave turbulence scaling law in $E(k_{\perp},k_{\parallel}) \sim 
 k_{\perp}^{-2}$. In the latter case  the parallel wavenumber is a mute variable because of the dynamical decoupling of parallel planes in Fourier space \cite{Galtier00}. 
 This prediction is given for the total energy, however, because in the small scale limit we have:
 \begin{equation}
 {\xi_{s}^{-s}}^{2}  \to \dfrac{1}{k^{2}d_{i}^{2}} \quad \rm{(L-polarity)} ,
 \end{equation}
and
\begin{equation}
 {\xi_{s}^{s}}^{2}  \to k^{2}d_{i}^{2} \quad \rm{(R-polarity)} ,
 \end{equation}
we can easily understand from equations (\ref{zab}) that either the magnetic or the velocity field will dominate in equation (\ref{heuristic_spectrum}) depending on which 
waves one considers.  Recalling that the magnetic field is tied in the electron flow, whereas the bulk velocity is carried by the heavy ions, one can understand that the 
emergence of the two circularly polarized waves with opposite polarity, is fundamentally linked to the microscopic scales with the opposite electric charge of the two species. 
These properties, which appear naturally if one projects the equations on a helical complex basis is completely hidden otherwise because Hall MHD is by construction a 
mono-fluid model.
\section{Kinematics}\label{sec3}
\subsection{Wave fluctuations}
In order to define the kinematics for the total energy of the 3D incompressible Hall MHD, we use, for symmetry reasons, the renormalized field $a_{\Lambda}^{s}$ defined by:
\begin{equation}
\mathcal{Z}_{\Lambda}^{s} = (\xi_{\Lambda}^{s}-\xi_{\Lambda}^{-s})a_{\Lambda}^{s}e^{-i\omega_{\Lambda}^{s}t}.
\label{ZversusA}
\end{equation}
We shall define statistical quantities by introducing the ensemble average denoted $\left\langle \cdot\cdot\cdot \right\rangle $. In practice 
for direct numerical simulations, we will use the ergodicity assumption and substitute it to a space integration 
(i.e. turbulence is homogeneous). We define:
\begin{equation}
E^{u}(\textbf{k})\equiv \sum_{\Lambda}\left\langle \mathcal{U}_{\Lambda}(\textbf{k})\mathcal{U}_{\Lambda}^{*}(\textbf{k})\right\rangle, 
\end{equation}
and
\begin{equation}
E^{b}(\textbf{k}) \equiv \sum_{\Lambda}\left\langle \mathcal{B}_{\Lambda}(\textbf{k})\mathcal{B}_{\Lambda}^{*}(\textbf{k})\right\rangle ,
\end{equation}
the kinetic and magnetic energy spectra respectively. The use of expressions (\ref{zab}) and (\ref{ZversusA}) leads to:
\begin{eqnarray}
\left\langle \vert \mathcal{U}_{\Lambda}\vert^{2} \right\rangle &=& {\xi_{\Lambda}^{-}}^{2}\left\langle \vert a_{\Lambda}^{+}\vert^{2} \right\rangle +  
{\xi_{\Lambda}^{+}}^{2}\left\langle \vert a_{\Lambda}^{-}\vert^{2} \right\rangle\\
&-& \left\langle a_{\Lambda}^{+}{a_{\Lambda}^{-}}^{*}e^{i(\omega_{\Lambda}^{+}-\omega_{\Lambda}^{-})t} \right\rangle -  
        \left\langle a_{\Lambda}^{-}{a_{\Lambda}^{+}}^{*}e^{i(\omega_{\Lambda}^{-}-\omega_{\Lambda}^{+})t} \right\rangle,\nonumber
\end{eqnarray}
and
\begin{eqnarray}
\left\langle \vert \mathcal{B}_{\Lambda}\vert^{2} \right\rangle &=& \left\langle \vert a_{\Lambda}^{+}\vert^{2} \right\rangle +  
                                                                                                           \left\langle \vert a_{\Lambda}^{-}\vert^{2} \right\rangle\\
&+& \left\langle a_{\Lambda}^{+}{a_{\Lambda}^{-}}^{*}e^{i(\omega_{\Lambda}^{+}-\omega_{\Lambda}^{-})t} \right\rangle +  
        \left\langle a_{\Lambda}^{-}{a_{\Lambda}^{+}}^{*}e^{i(\omega_{\Lambda}^{-}-\omega_{\Lambda}^{+})t} \right\rangle.\nonumber
\end{eqnarray}
Note that the Hermitian symmetry property of real-valued quantity  (conjugate symmetry) is used for the term in the right hand side. 
Then, the total energy spectrum becomes:
\begin{eqnarray}
E^{u}(\textbf{k})+E^{b}(\textbf{k}) &=& ({\xi_{+}^{-}}^{2}+1)\left\langle \vert a_{+}^{+}\vert^{2} \right\rangle\\
+({\xi_{-}^{+}}^{2}+1)\left\langle \vert a_{-}^{-}\vert^{2} \right\rangle &+& ({\xi_{+}^{+}}^{2}+1)\left\langle \vert a_{+}^{-}\vert^{2} \right\rangle\nonumber\\
&+& ({\xi_{-}^{-}}^{2}+1)\left\langle \vert a_{-}^{+}\vert^{2} \right\rangle.\nonumber
\end{eqnarray}
These results lead us to define the left and right polarized kinetic and magnetic energies as: 
\begin{eqnarray}
E^{u}_{L}(\textbf{k})&\equiv & {\xi_{+}^{+}}^2\left\langle \vert a_{+}^{-}\vert^2\right\rangle + {\xi_{-}^{-}}^2\left\langle \vert a_{-}^{+}\vert^2\right\rangle,
\\
E^{u}_{R}(\textbf{k})&\equiv & {\xi_{+}^{-}}^2\left\langle \vert a_{+}^{+}\vert^2\right\rangle + {\xi_{-}^{+}}^2\left\langle \vert a_{-}^{-}\vert^2\right\rangle,\\
E^{b}_{L}(\textbf{k})&\equiv & \left\langle \vert a_{+}^{-}\vert^2\right\rangle +\left\langle \vert a_{-}^{+}\vert^2\right\rangle,\\
E^{b}_{R}(\textbf{k})&\equiv &\left\langle \vert a_{+}^{+}\vert^2\right\rangle +\left\langle \vert a_{}^{-}\vert^2\right\rangle.
\end{eqnarray}
These definitions provide two important relationships at the level of kinematics:
\begin{equation}
E^{u}_{L}(\textbf{k}) = {\xi_{+}^{+}}^2 E^{b}_{L}(\textbf{k}) =  {\xi_{-}^{-}}^2 E^{b}_{L}(\textbf{k}),
\label{r1}
\end{equation}
and 
\begin{equation}
E^{b}_{R}(\textbf{k}) = {\xi_{+}^{-}}^2 E^{u}_{R}(\textbf{k}) = {\xi_{-}^{+}}^2 E^{u}_{R}(\textbf{k}).
\label{r2}
\end{equation}
Expressions (\ref{r1}) and (\ref{r2}) tell us that in the small scale limit $k d_{i} \gg 1$ the dynamics is mainly driven by the velocity field for the L fluctuations and by the 
magnetic field for the R fluctuations. 
Knowing $E^{u}_{L}(\textbf{k})$ or $E^{b}_{R}(\textbf{k})$ automatically gives the form of the spectra for the corresponding fluctuations with the same 
type of polarity, the latter being driven by the former. Remarkably this scenario is also applicable in the isotropic case as shown in \cite{Meyrand12}, for which the 
concept of polarization was generalized. 
\subsection{Two-dimensional state}
In the different relationships derived above, terms of mixed polarities ($\propto e^{i(\omega_\Lambda^s - \omega_\Lambda^{-s})t}$) appear. This type of contribution is 
expected to be weaker than the others (pure real terms) because the presence of a mean magnetic field leads dynamically to the separation between the time-scales of 
amplitudes and phases with, thus, a tendency to a phase mixing. It is in the weak turbulence limit that the phase mixing is the strongest: in this case these contributions 
tend asymptotically to zero. The situation is, however, different for the two-dimensional state which corresponds by definition to $k_{\parallel}=0$. For this state we may 
simplify the terms of the mixed polarities because we have $\omega_\Lambda^s=0$. The phase mixing does not operate at all and a significant contribution to the 
kinematics of the mixed polarity terms may be expected. 
\section{Spectral Properties: anisotropy, scaling, transfer and flux}\label{sec4} 
\subsection{Simulation setup} 
We solve the incompressible Hall MHD equations in a periodic, rectangular domain with aspect ratio $L_{\perp}^{2} \times L_{\parallel}$ using the 
TURBO code \cite{Teaca09}, in which we have implemented the Hall term. A series of benchmarks including those against exact non-linear solutions  \cite{Mahajan05} of the 
Hall MHD equations are presented in \cite{Meyrand_Thesis}.  We set $L_{\perp}=L_{\parallel}=2\pi$ , $d_{i}=0.5$ and $b_{0}=25$. A 3D pseudospectral algorithm is used 
to perform the spatial discretization on a grid with a resolution of $N_{\perp}^{2}\times N_{\parallel}$ mesh points (see Table \ref{table:tablo}). 
The time step is computed automatically by a 
Courant-Friedricks-Lewey $\text{CFL}=0.3$ criterion and the time advancement is based on a modified Williamson, four-step, third-order low storage Runge- Kutta 
method \cite{Williamson80}. To save on computational costs,  we have reduced the field-parallel numerical resolution with $N_{\perp} >N_{\parallel}$. This is appropriate 
since the energy cascade proceeds much faster in the field-perpendicular direction, resulting in an anisotropy in $\textbf{k}$-space such that the energy at large 
$k_{\parallel}$ is reduced.

The initial state consists of isotropic magnetic and velocity field fluctuations with random phases such that the total cross-helicity $H^{c}=\left\langle 
\textbf{u}\cdot\textbf{b}\right\rangle $, as well as the total magnetic helicity $H^{m}=\left\langle \textbf{a}\cdot\textbf{b}\right\rangle$ and kinetic helicity 
$H^{k}=\left\langle \textbf{u}\cdot\nabla\times\textbf{u} \right\rangle$ are zero ($\left\langle \cdot\cdot\cdot \right\rangle $ denotes a volume average). The initial 
kinetic and magnetic energies are equal to $1/2$ and localized at the largest scales of the system (wave numbers $\vert\textbf{k}\vert d_{i}\in [1,2]$ are initially 
excited).

\begin{table}
\begin{tabular}{| c | c  c  c  c |}
\hline 
\hline 
Run & $~~~~N_{\perp}~~~~$ & $~~~~N_{\parallel}~~~~$ & $~~~~\nu_{3}~~~~$ & $~~~~\eta_{3}~~~~$ \\ 
\hline 
\nbRoman{1} Hall MHD & 768 & 512 & $5\times 10^{-13}$ & $2\times 10^{-11}$ \\ 

\nbRoman{2} Hall MHD & 768 & 256 & $1.7\times 10^{-3}$ & $2\times 10^{-11}$ \\ 

\nbRoman{3} Hall MHD & 128 & 64 & $1.7\times 10^{-9}$ & $6.4\times 10^{-9}$ \\ 

\nbRoman{4} Hall MHD & 256 & 128 & $7.6\times 10^{-11}$ & $4.5\times 10^{-10}$ \\ 

\nbRoman{5} EMHD & 256 & 128 & $\varnothing$ & $4.5\times 10^{-10}$ \\
\hline 
\hline
\end{tabular} 
\caption{Summary of the simulations parameters.}
\label{table:tablo} 
\end{table}

\begin{figure}
\includegraphics[width=1.0\linewidth]{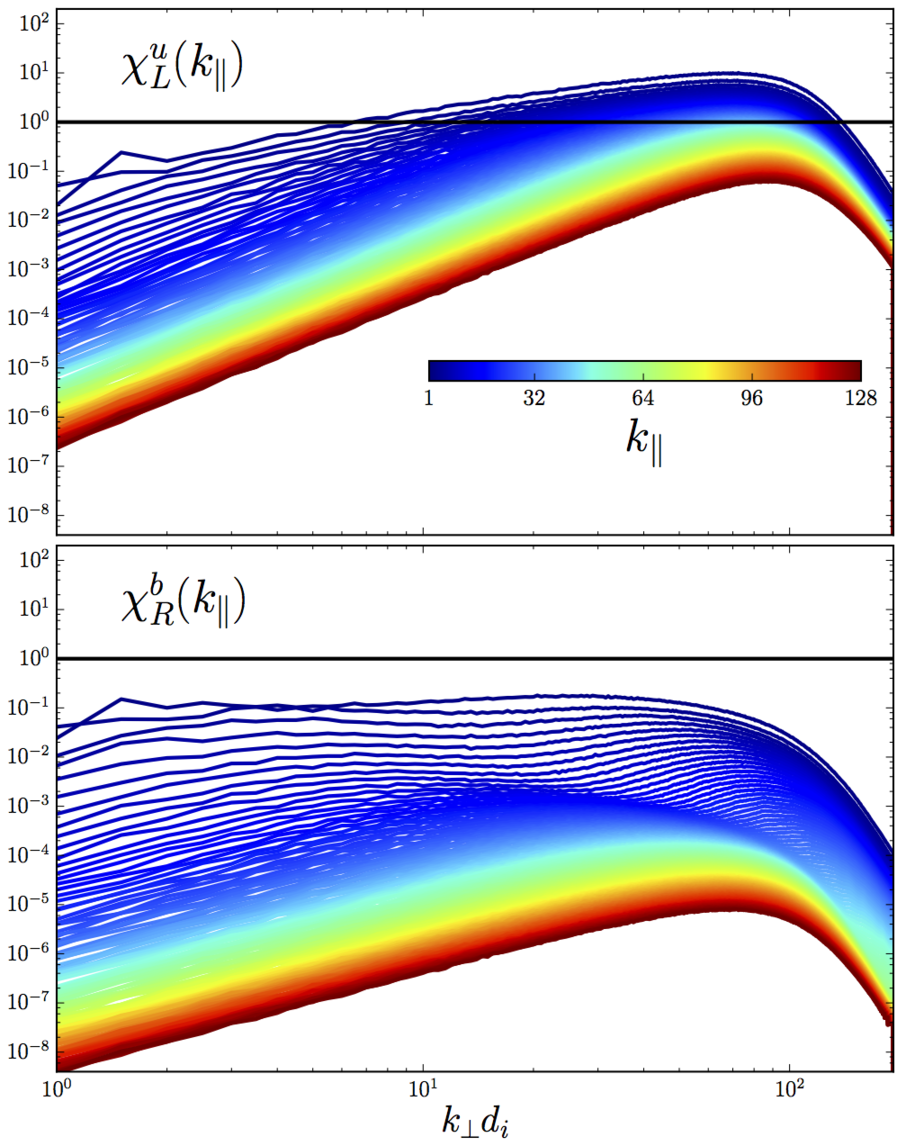}
\caption{Transverse wave number dependence of the time ratios $\chi_{L}^{u}$ (top) and $\chi_{R}^{b}$ (bottom) for different values of $k_{\parallel}$ at 
$t \sim 90\omega_{ci}$ (Run \nbRoman{1}). The horizontal line marks the demarcation between weak (below) and strong (above) wave turbulence.}
\label{fig:chi}
\end{figure} 
\begin{figure*}
      \subfigure[]{\label{fig:spectrum_u}}\includegraphics[width=0.45\textwidth]{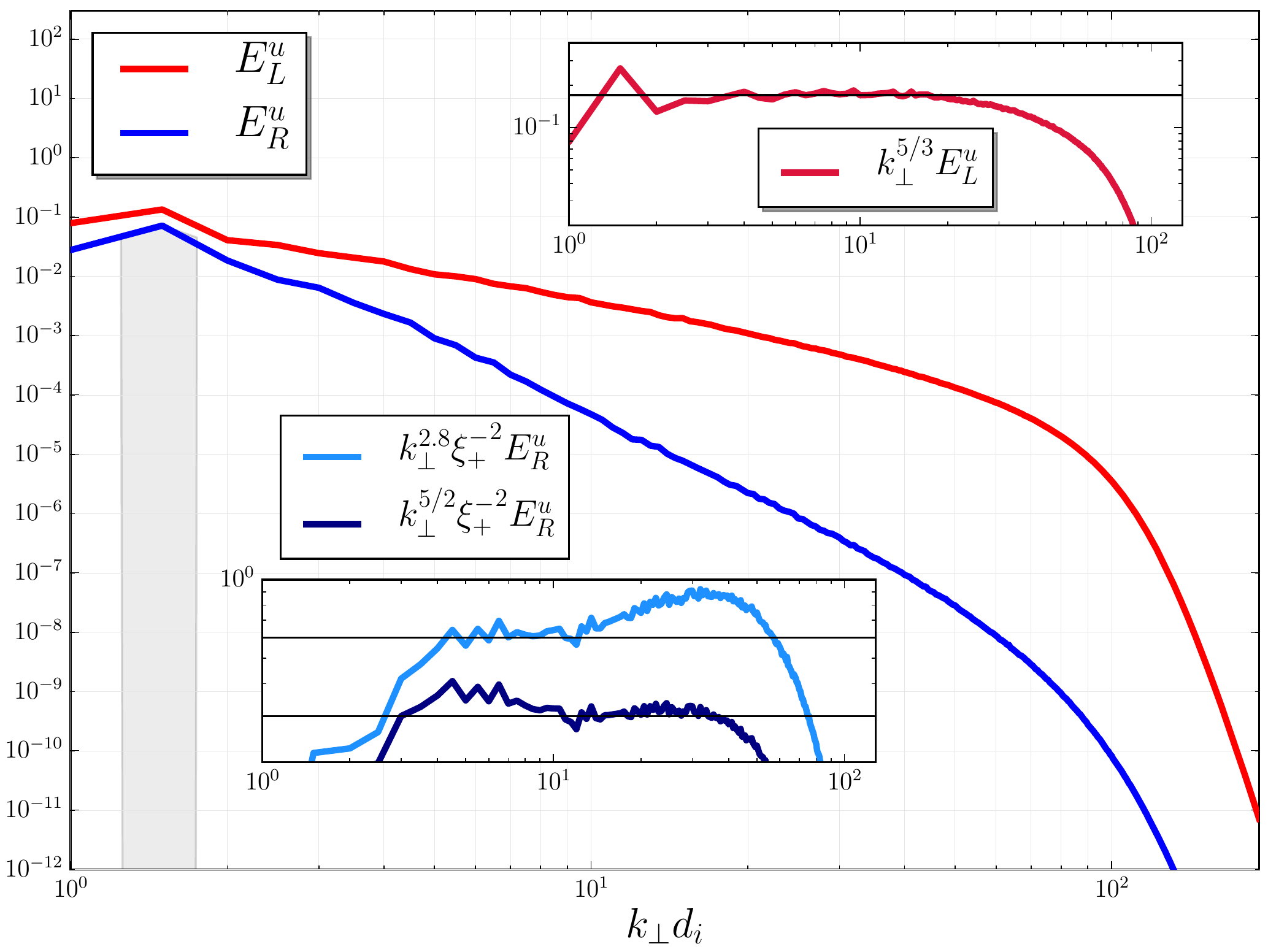}   
      \subfigure[]{\label{fig:spectrum_b}} \includegraphics[width=0.45\textwidth]{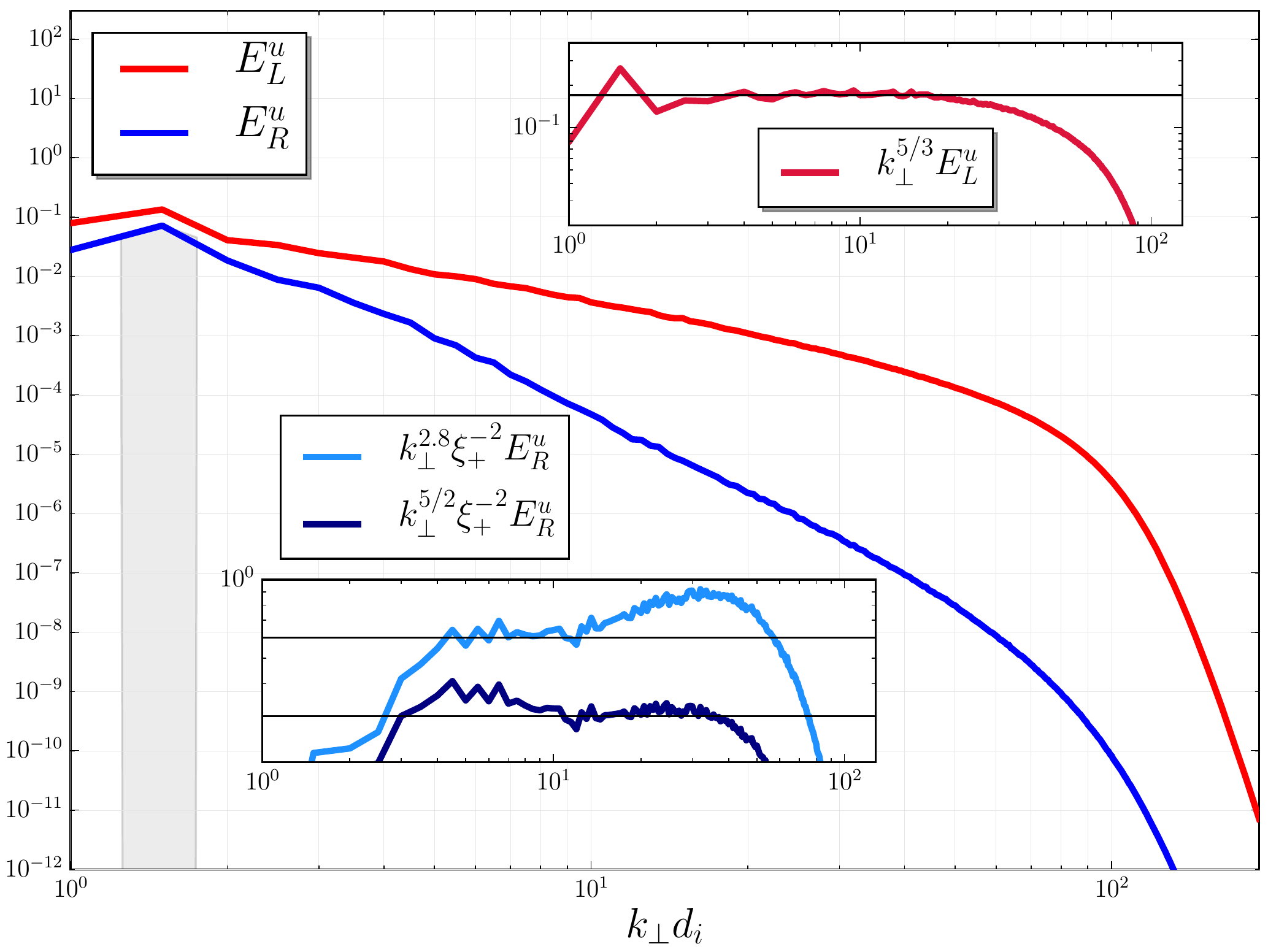}
    \caption{\subref{fig:spectrum_u}: Transverse velocity spectra of the L-fluctuations (red) and R-fluctuations (blue) (Run \nbRoman{1}). The grey column corresponds 
    to the forcing scales and the inserts shows L and R compensated spectra. \subref{fig:spectrum_b}: Same as Fig. \ref{fig:spectrum_u} but for the magnetic L (red) and R 
    (blue) fluctuations spectra.}
    \label{spectra}
\end{figure*}
For the purposes of this study we have developed a helical forcing. The turbulence is driven at the largest scales with a forcing local in Fourier space. It acts on all the modes 
within the shell $s_{f}$ defined by $\vert \textbf{k} \vert d_{i} \in \left[1.25, 1.75\right]$. In practice, the kinetic $\hat{\textbf{f}}_{u}$ and magnetic $\hat{\textbf{f}}_{b}$ 
forces have the form
\begin{eqnarray}
\hat{\textbf{f}}_{u}(\textbf{k}) =  \sum_{\Lambda} \alpha_{\Lambda}(\textbf{k}) \mathcal{U}_{\Lambda}(\textbf{k}) \textbf{h}_{k}^{\Lambda},\\
\hat{\textbf{f}}_{b}(\textbf{k}) =  \sum_{\Lambda} \beta_{\Lambda}(\textbf{k}) \mathcal{B}_{\Lambda}(\textbf{k}) \textbf{h}_{k}^{\Lambda} .
\end{eqnarray}
The parameters $\alpha_{\Lambda}(\textbf{k})$ and $\beta_{\Lambda}(\textbf{k})$ are given by 
\begin{eqnarray}
\alpha_{\Lambda}(\textbf{k}) = \dfrac{\varepsilon^{u}_{\Lambda}}{E^{u}_{\Lambda}(\textbf{k}) N_{f}^{2}},\\
\beta_{\Lambda}(\textbf{k}) = \dfrac{\varepsilon^{b}_{\Lambda}}{E^{b}_{\Lambda}(\textbf{k}) N_{f}^{2}},
\end{eqnarray}
where $2E^{u}_{\Lambda}(\textbf{k})=\vert\mathcal{U}_{\Lambda}(\textbf{k})\vert^2$, $2E^{b}_{\Lambda}(\textbf{k})=\vert\mathcal{B}_{\Lambda}(\textbf{k})\vert^{2}$  
and $N_{f}$ is the number of forced modes. This choice ensures that each of the $N_{f}$ forced modes is submitted to a forcing mechanism that injects kinetic and 
magnetic energy at the constant rates $\sum_{\Lambda} \varepsilon^{u}_{\Lambda}$ and $\sum_{\Lambda} \varepsilon^{b}_{\Lambda}$. We fix
$\varepsilon^{u}_{+}=\varepsilon^{u}_{-}=\varepsilon^{b}_{+}=\varepsilon^{b}_{-}= 0.025$ which enforces the kinetic helicity and magnetic helicity injection rates to be zero. 
Remarkably this choice turns out to impose a cross-helicity level close to zero. 
Note that in the momentum equation (\ref{hmhd2}), the kinetic forcing 
$\textbf{f}^u$, whatever its precise form, can always be considered as divergence free since the pressure will enforce the incompressibility of the velocity field by 
eliminating any $\nabla\cdot\textbf{f}^{u}$ contribution of the force. 
On the other hand, $\textbf{f}^{b}$ must always be divergence free as a consistency condition for the magnetic field. 
This latter condition is automatically satisfied by the helical nature of the forcing. Importantly, since the forces are proportional to the fields, the 
characteristic time of the forces will tend to be equal to that of the intrinsic characteristic time of the large scale eddies (corresponding to modes within the shell $s_{f}$). 
With such a forcing we therefore do not introduce any artificial and potentially dynamically disturbing characteristic times. Furthermore, since the $\alpha$'s and 
$\beta$'s parameters are real, the forcing method presented here does not influence the phases of the fields,  which ensures that no change is made in the type of 
turbulent structures present. 

The system is evolved until a stationary state is reached for both the velocity and the magnetic fields, which is confirmed by observing the time evolution of the total energy 
as well as the dissipation rate of total energy of the fluctuations (not shown). Note that, in order to achieve a stationary state, it is necessary to remove the amount of ideal invariants that may be injected to the system by the forcing mechanism. In order to achieve this, we used kinetic hyper-dissipation $\nu_{3}\Delta^{3}$ and magnetic hyper-diffusivity $\eta_{3}\Delta^{3}$.
When the stationary state is reached, the kinetic and magnetic energy fluxes relax to a level constrained by the kinetic hyper-dissipation and magnetic hyper-diffusivity respectively. 

As shown in Table \ref{table:tablo}, we conducted a number of runs to investigate various aspects of 3D incompressible Hall MHD turbulence.

\subsection{Domain of validity of WTT}\label{section:domain_of_validity}
WTT deals with asymptotic developments which are based on a time scale separation, with a non-linear time assumed to be much larger than the wave period. Consequently, 
a necessary condition for the existence of weak turbulence is that the ratio between non-linear and linear time-scales is small compare to one. We shall therefore evaluate 
the turbulence regime by considering the different time-scales of the problem.  In Hall MHD because two  waves with two different dispersion relations exist at sub-ion 
scales, it is necessary to define two different non-linear time-scales.  Because the left handed ion cyclotron waves are associated with the velocity field (see discussion in 
Section \ref{heuristic}) we may define the corresponding left handed non-linear time-scale from the momentum equation (\ref{hmhd2}) as $\tau_{nl}^{L} \sim 
1/(k_{\perp}u_{L})$. \textit{A contrario} because the right handed whistler waves are associated with the magnetic field, we may define the right handed non-linear 
time-scale from the  Maxwell-Faraday's equation (\ref{hmhd3}) as $\tau_{nl}^{R} \sim 1/(d_{i}k_{\perp}^{2}b_{R})$. Therefore, the asymptotic condition $\tau_{tr} \gg \tau_{w}$ 
of WTT implies that the following two relations are fulfilled simultaneously: 
\begin{equation}
\left\{
      \begin{aligned}
\chi_{L}^{u} = \dfrac{\tau_{ci}}{\tau_{nl}^{L}} \sim \dfrac{d_{i}k_{\perp}^2 u_{L}}{k_{\parallel}b_{0}}\ll 1, \label{chi}\\
\chi_{R}^{b} = \dfrac{\tau_{w}}{\tau_{nl}^{R}} \sim \dfrac{k_{\perp}b_{R}}{k_{\parallel}b_{0}}\ll 1.
 \end{aligned}
    \right.
\end{equation}
\begin{figure*}
       {\label{fig:spectrum_ul2D}}\includegraphics[width=1.0\linewidth]{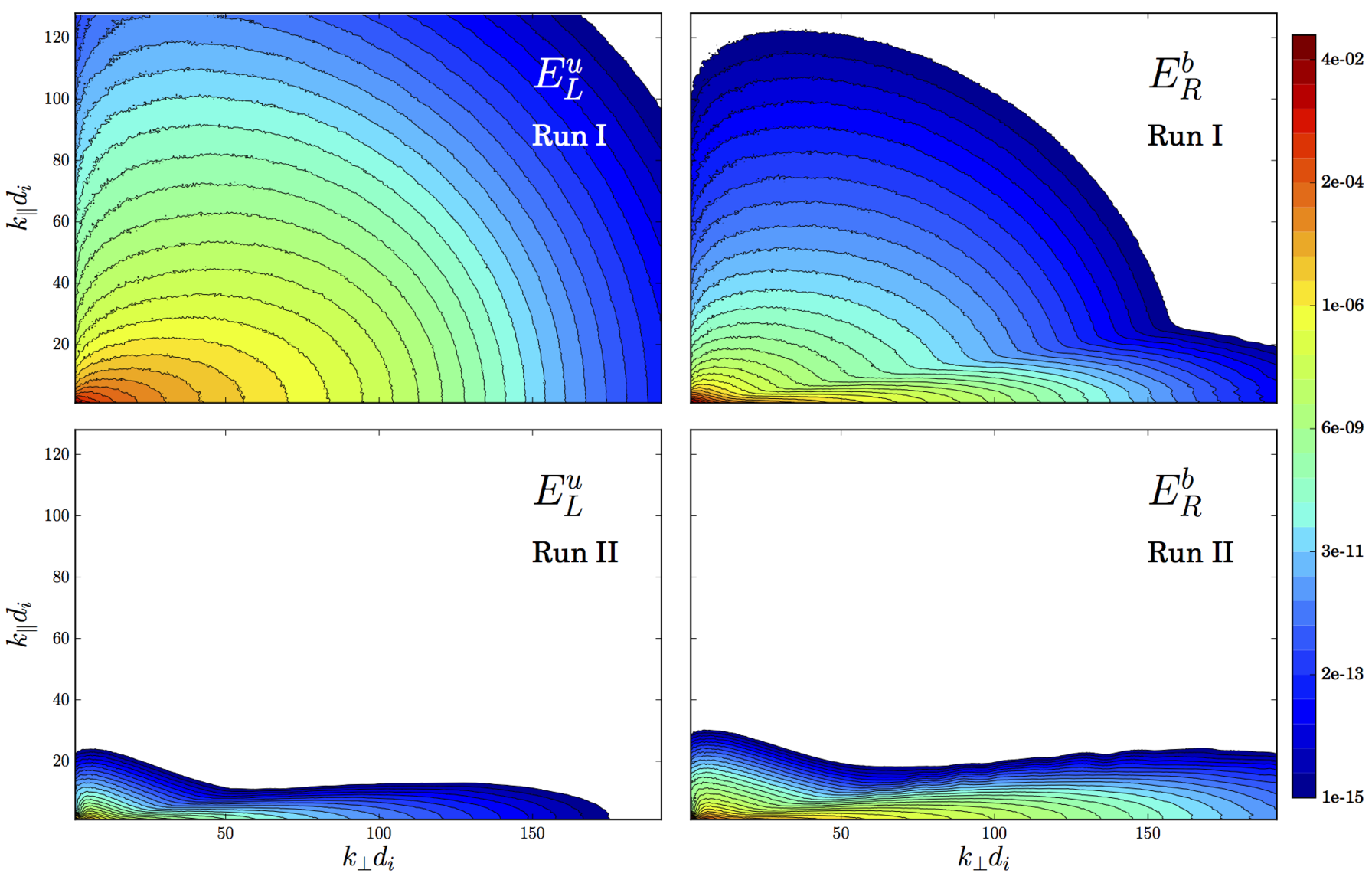}
        \caption{Iso-contours (in logarithmic scale) of the bi-dimensional L velocity energy spectrum $E^{u}_{L}(k_{\perp},k_{\parallel})$ (left) and R magnetic energy 
        spectrum $E^{b}_{R}(k_{\perp},k_{\parallel})$ (right) corresponding to Run \nbRoman{1} (top) and \nbRoman{2} (bottom).}  
    \label{fig:spectra2D}  
\end{figure*}
It is clear from these two equations that the WTT is not uniformly valid in all of $\textbf{k}$-space and that its range of applicability can be different for L and R turbulent 
fluctuations. If one substitutes the WTT predictions for the kinetic and magnetic energy spectra in (\ref{chi}), we can estimate the $k_{\perp}$ dependence of $\chi_{L}^{u}$ 
and $\chi_{R}^{b}$. It gives, 
\begin{equation}
 \left\{
      \begin{aligned}
\chi_{L}^{u} &\propto k_{\perp}^{5/4}, \label{chi2}\\
\chi_{R}^{b} &\propto k_{\perp}^{1/4}. 
 \end{aligned}
    \right.
\end{equation}
This means that the degree of non-linearity of left handed ion cyclotron fluctuations increases much more rapidly than the degree of non-linearity of right handed whistler 
fluctuations. This situation has a profound impact on the turbulent dynamics as we will see below.

The plots of $\chi_{L}^{u}$ and $\chi_{R}^{b}$ corresponding to Run I are given in Fig.\,\ref{fig:chi} for different values of $k_{\parallel}$. For this evaluation, $u_{L}$ 
and $b_{R}$ are respectively defined as $b_{R} = \sqrt{2k_{\parallel}k_{\perp}E^{b}_{R}(k_{\parallel},k_{\perp})}$ and $u_{L} = \sqrt{2k_{\parallel}k_{\perp}E^{u}_{L}
(k_{\parallel},k_{\perp})}$. The axisymmetric bi-dimensional magnetic and kinetic spectra $E^{b,u}(k_{\parallel},k_{\perp})$ are linked to the magnetic and kinetic 
energies ${\cal E}^{b,u}$ of the system through the relation ${\cal E}^{b,u} = \iint E^{b,u}(k_{\parallel},k_{\perp})dk_{\perp}dk_{\parallel}$. 
We clearly see that the R fluctuations belong to the weak turbulence regime for all $k_{\perp}$ and $k_{\parallel} > 0$. 
For the L fluctuations the situation is radically different. There exists a 
critical scale around $k_{\perp}d_{i} \sim 6$ beyond which the weak turbulence cascade drives itself into a state which no longer satisfies the premise on which the 
theory is based. Note that a similar situation is excepted in Alfv\'en wave turbulence and has been observed in direct numerical simulations \cite{Meyrand16}.  If it is 
true that some L modes belong to the weak turbulence regime for all $k_{\perp}$  (those for which $k_{\parallel} > 64$), they actually do not contribute significantly to 
the dynamics because they are energetically sub-dominant by several orders of magnitude.
\begin{figure*}
 \centering
    \subfigure[]{\label{fig:w_perp}\includegraphics[width=0.49\textwidth]{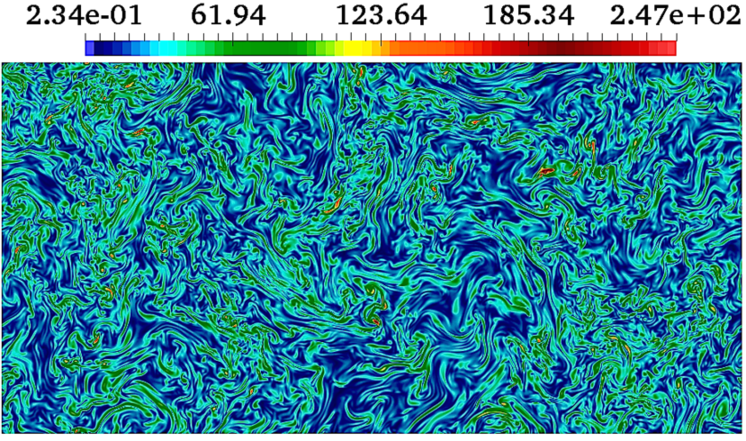}}
    \subfigure[]{\label{fig:j_perp}\includegraphics[width=0.49\textwidth]{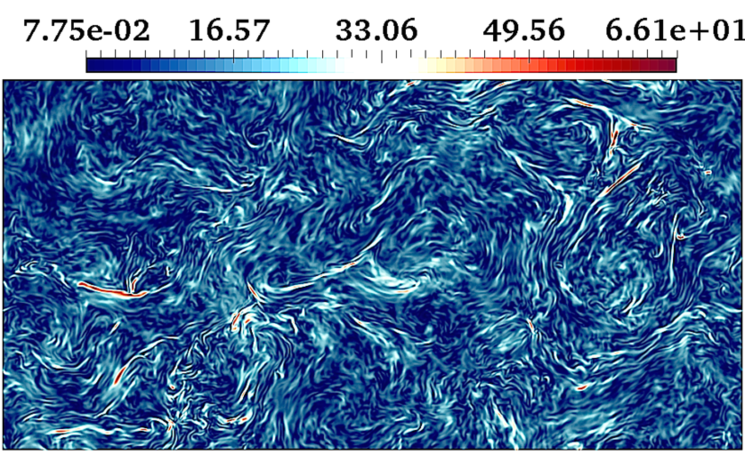}}
    \subfigure[]{\label{fig:w_para}\includegraphics[width=0.49\textwidth]{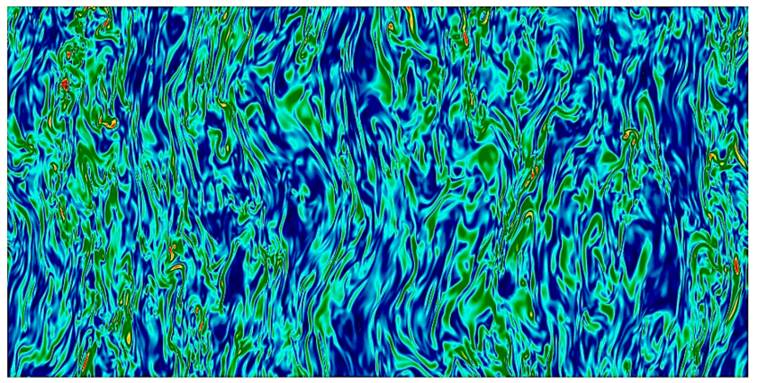}}
    \subfigure[]{\label{fig:j_para}\includegraphics[width=0.49\textwidth]{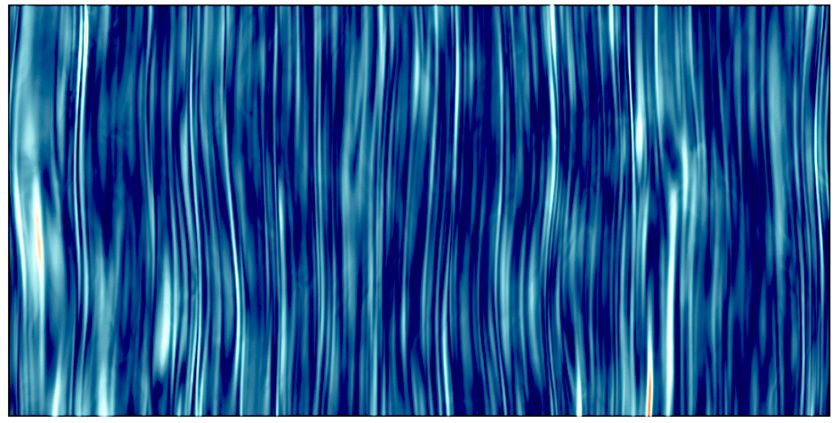}}
    \caption{Amplitude of vorticity (left) and current density (right) fluctuations in a field-perpendicular (top) and field-parallel (bottom) cross section of the 
simulation domain (Run I). Clearly the anisotropy is stronger for the current density than the vorticity fluctuations. The width of the snapshots is equal to $2\times d_{i}$ 
and the height to $1\times d_{i}$.}
    \label{fig:real-space-fields}
\end{figure*}  

The present numerical simulation consists therefore in a non-trivial superposition of mainly highly non-linear ion cyclotron modes and weakly non-linear whistler waves. 
This situation may seem peculiar and anecdotal corresponding to a very specific parameter regime. One may think for example that by increasing the strength of the 
mean magnetic field and/or by diminishing the amplitude of the forcing one may reach a pure weak turbulence regime for both L and R fluctuations. 
But the situation proves to be more subtle because the linear term for the velocity field (mainly L fluctuations) involves the magnetic field (mainly R fluctuations) (see 
equation (\ref{hmhd2})).  As one increases the strength of the mean magnetic field ${\bf b_0}$, the R energy cascade weakens while \textit{a contrario} the L energy 
cascade, which is always less weak (see equations (\ref{chi2})), intensifies accordingly as the total energy injection rate is constant (see discussion in section \ref{section:transfers}). 
Thereby, the relative importance of the linear term with respect to the non-linear term in the momentum equation (\ref{hmhd2}) is not proportional to $b_{0}$. 
The superposition of highly non-linear ion cyclotron modes and weakly non-linear whistler waves appears therefore to be a standard situation in Hall MHD turbulence. 
This ascertainment raises fundamental questions about the applicability of WTT to Hall MHD. The first step to address this question is to verify if some of the Hall 
MHD WTT predictions can be recovered from these numerical experiments. 
\subsection{Properties of the energy spectra}\label{section:spectra}
Figure \ref{spectra} displays the one-dimensional axisymmetric transverse velocity and magnetic spectra (an integration over a cylinder whose axis of symmetry is 
$\textbf{b}_{0}$ is made) for the L and R  fluctuations corresponding to Run \nbRoman{1} at time $t\sim 90 \omega_{ci}$ for which the simulation is statistical 
stationary. As predicted by the kinematics, there is a ${\xi_{\Lambda}^{-s}}^2$ of difference between the spectra of same polarity. Consequently the magnetic energy is 
dominated by R fluctuations whereas the kinetic energy is dominated by L fluctuations.  The velocity field follows a Kolmogorovian spectrum in $k_{\perp}^{-5/3}$ while the 
magnetic spectrum presents a knee around $ k_{\perp}d_{i}\sim 10$ with a change in slope going from approximately $k_{\perp}^{-2.8}$ to the WTT prediction in 
$k_{\perp}^{-5/2}$.

Figure \ref{fig:spectra2D} (top-left) displays the iso-contours of the bidimensional L--kinetic energy spectrum $E^{u}_{L}(k_{\perp},k_{\parallel})$ corresponding to Run 
\nbRoman{1}. At large-scale ($k<50$) the iso-contours are elongated along the $k_{\perp}$ direction which can be interpreted  as a direct consequence of the 
weak ion-cyclotron wave dynamics.  At smaller scales, one can observe a progressive stretching of the iso-contours in the $k_{\parallel}$ direction which is due to the 
transition toward strong ion-cyclotron wave turbulence as expected from the spectral properties of the $\chi^{u}_{L}$ parameter (see Fig. \ref{fig:chi}).
Figure \ref{fig:spectra2D} (top-right) displays the iso-contours of the bidimensional R--magnetic $E^{b}_{R}(k_{\perp},k_{\parallel})$ energy spectrum for the same simulation. 
Interestingly, one can observe the presence of two lobes: one with a strong anisotropy in the $k_{\perp}$ direction and another which extends in the $k_{\parallel}$ 
direction showing a propensity toward isotropization as $k_{\parallel}$ increases. This later property is similar to the one observed for the bi-dimensional L velocity energy 
spectrum (Fig. \ref{fig:spectra2D} top-left) and suggests a possible coupling between ion cyclotron and whistler waves. The difference in anisotropy between the velocity and 
magnetic field is clearly visible in real space as can be seen in Fig. \ref{fig:real-space-fields}. We clearly see that the amplitude of the current density fluctuations are more 
elongated along the vertical (i.e. ${\bf b_0}$) direction than the vorticity. 
Several direct numerical experiments provide convincing evidence that electron MHD (EMHD) turbulence develops a strong anisotropy in the presence of a mean magnetic 
field \cite{Cho04, Ng03, Cho09, Meyrand13}. Yet, EMHD corresponds to Hall MHD with $\textbf{u} = \textbf{0}$, the difference between Run \nbRoman{1} and those 
previously cited may therefore be due to the velocity dynamics. Note, however, that despite the fact that anisotropy of EMHD turbulence is usually considered as granted 
and turns out to be mandatory to justify the use of gyrokinetics to model solar wind turbulence \cite{Schekochihin09, Sahraoui2012}, whistler 
resonant three-wave interactions allow energy transfer along the external magnetic field. This properties contrasts with the Alfv\'en resonant triads 
which foliate the wavevector space, a property which strictly forbids 
any parallel cascade. In weak EMHD turbulence the transfer along the mean field direction is small solely if local interactions in $k_{\perp}$ are dominants 
\cite{Dastgeer00}. In this case only counter-propagative whistler waves contribute significantly to the non-linear dynamics \cite{Galtier03} and small scales are preferentially 
generated perpendicular to the external magnetic field. The presence of the lobe of energy in the parallel direction may therefore also be due to non-local interactions. 

A straightforward way to discriminate this two scenarios (non-local {\it versus} ion-cyclotron/whistler interactions) is to perform a numerical experiment with a hyper-viscosity 
large enough to act at large scale. \textit{A priori} in this circumstance the L-fluctuations cannot develop their own non-linear dynamics. Run \nbRoman{2} described in 
Table \ref{table:tablo} corresponds precisely to this situation. Figure \ref{fig:spectra2D} (bottom-right) displays the iso-contours of the bidimensional R-magnetic energy 
spectrum for this simulation. Clearly, the extent of lobe in the $k_{\parallel}$ direction is greatly reduced which strongly suggests  that the anomalous spectrum observed 
in Run \nbRoman{1} is due to a cross-coupling between ion-cyclotron and whistler waves. 
The iso-contours of the bi-dimensional L-velocity energy spectrum displayed in Fig. \ref{fig:spectra2D} (bottom-left) are homothetic to the R-magnetic one under a 
$\xi^{+}_{-}$ transformation. Interestingly the velocity spectrum extends at scales much smaller than the Kolmogorov dissipation micro-scale. This property reflects the 
fact that the velocity is enslaved to the magnetic field {\it via} the Lorentz force term $\textbf{j}\times\textbf{b}$ which is not directly affected by the velocity hyper-diffusive 
term.  The presence of a small lobe in the $k_{\parallel}$ shows, however, a small back reaction of the velocity {\it via} the non-linear advection term 
$\textbf{u}\cdot\nabla\textbf{u}$. This demonstrates that contrary to a common believe, EMHD is not simply the 
small scale limit $kd_{i}\gg 1$ of Hall MHD and that, whatever the value of the Prandtl number \cite{Meyrand12}. 

To investigate further the cross-coupling between ion-cyclotron and whistler waves we perform in section \ref{section:transfers} a detailed analysis of the shell-to-shell 
energy transfer functions.  
\subsection{Shell-to-shell energy transfer functions}\label{section:transfers}
\begin{figure}
\includegraphics[width=\linewidth]{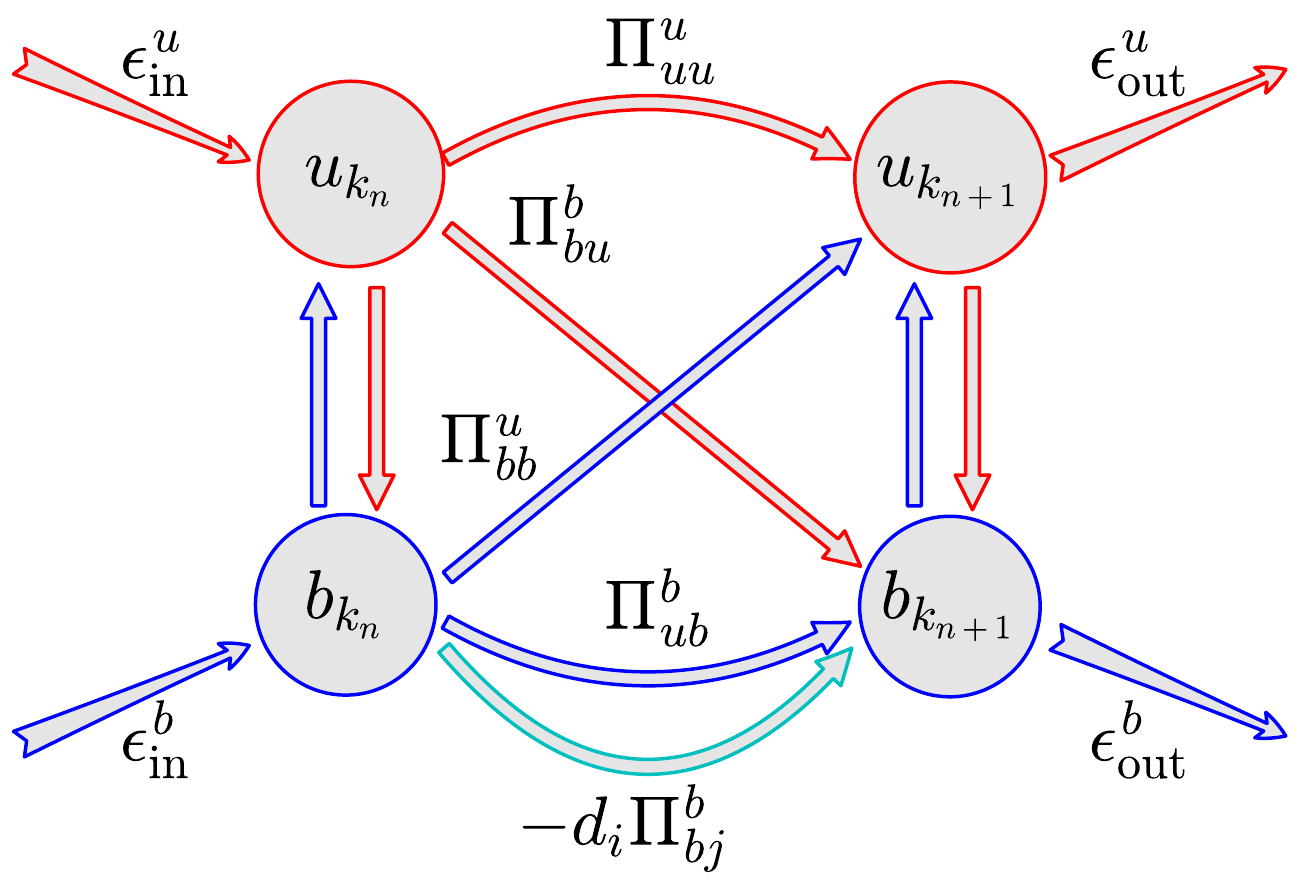}
         \caption{Schematic representation of various energy fluxes at a given scale in the inertial range of Hall MHD turbulence. 
         Note that because kinetic and magnetic energies are not conserved separately, non-trivial energy fluxes inside a same wave vector domain may exist (vertical arrows). 
         The arrows labelled $\epsilon^{u,b}_{\text{in,out}}$ represent the different incoming/outgoing energy fluxes.}
         \label{fig:flux_schema}
\end{figure}
The shell-to-shell kinetic and magnetic energy transfer functions are defined by: 
 \begin{eqnarray}
& \dfrac{\partial E^{u}(\textbf{k})}{\partial t} =\displaystyle\sum_{\textbf{p}}\left[ T^{u}_{uu}(\textbf{k},\textbf{p})-T^{u}_{bb}(\textbf{k},\textbf{p})\right] ,\label{dtEu}\\
&\dfrac{\partial E^{b}(\textbf{k})}{\partial t} =\displaystyle\sum_{\textbf{p}}\left[ T^{b}_{bu}(\textbf{k},\textbf{p})-T^{b}_{ub}(\textbf{k},\textbf{p})-d_{i}T^{b}_{bj}(\textbf{k},
\textbf{p})\right],\quad\quad
\label{dtEb}
 \end{eqnarray}
where 
\begin{equation}
T^{X}_{YZ}(\textbf{k},\textbf{p}) = \sum_{\textbf{q}}\mathfrak{Im}\lbrace[\textbf{k}\cdot \hat{\textbf{Z}}(\textbf{p}) ][\hat{\textbf{Y}}(\textbf{q})\cdot \hat{\textbf{X}}^{*}
(\textbf{k})]\rbrace \delta_{\textbf{q}+\textbf{p},\textbf{k}},
\label{transfers}
\end{equation}
is the transfer function to the mode $\textbf{k}$ of field $\textbf{X}$ from mode $\textbf{p}$ of field $\textbf{Z}$, mediated by all possible triadic interactions with 
modes $\textbf{q}$ of fields $\textbf{Y}$ that respects the condition $\textbf{k}=\textbf{p}+\textbf{q}$. $\mathfrak{Im}$ denotes the imaginary part 
and the asterisk the complex conjugate. 
Note that for the sake of clarity we omit the hyper-dissipative $T^{u}_{diss.} = 2\nu_{3}k^{6}E^{u}(\textbf{k})$ and hyper-diffusive 
$T^{b}_{diss.} = 2\eta_{3}k^{6}E^{b}(\textbf{k})$ terms as well as the forcing terms. 
The energy flux flowing toward a given $k$--scale {\it via} the $T^{X}_{YZ}$ channel is given by
\begin{equation}
\Pi_{YZ}^{X}(\textbf{k}) = \sum_{\textbf{k}^{'}=0}^{\textbf{k}} T^{X}_{YZ}(\textbf{k}^{'},\textbf{p}).
\label{flux}
\end{equation}
To study the perpendicular cascade, we consider concentric cylindrical shells along $\textbf{b}_0$ with constant width on a logarithmic scale which we define as the 
region $k_{0}2^{n/4}\leq k_{\perp}d_{i} \leq k_{0}2^{(n+1)/4}$ for the shells numbered $4\leq n\leq N$, where we set $k_{0}=2$ and $N=25$. A schematic representation 
of the various energy fluxes that we may find {\it a priori} in the inertial range of Hall MHD turbulence is given in Fig. \ref{fig:flux_schema}.
Since the kinetic and magnetic energies are not inviscid and ideal invariants, non-trivial energy fluxes may also exist between them at a given scale.
\begin{figure}
\includegraphics[width=\linewidth]{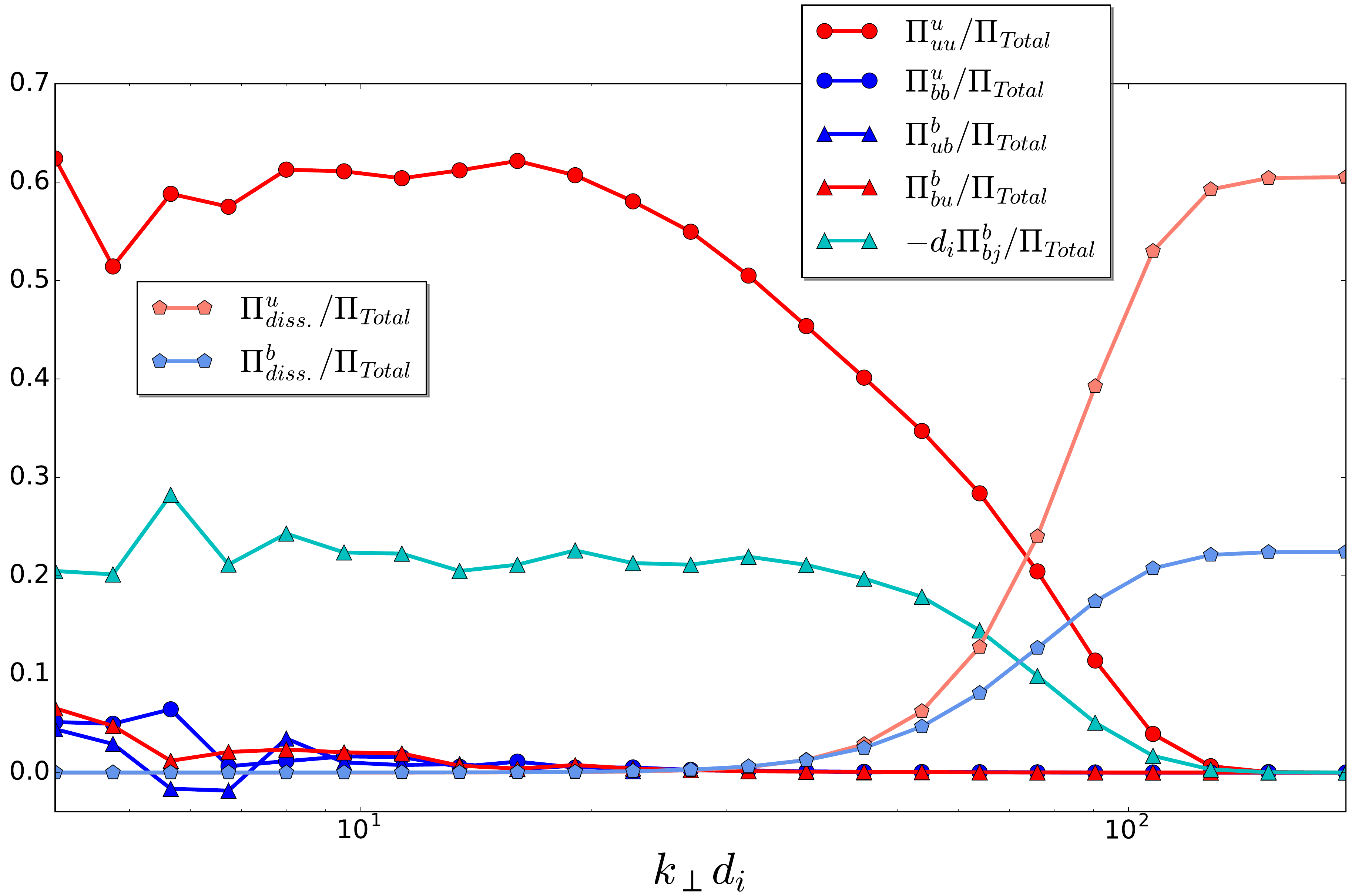}
         \caption{Normalized energy fluxes {\it versus} $k_\perp d_{i}$ (Run \nbRoman{1}).}
         \label{fig:flux}
\end{figure}

The various energy fluxes normalized by the total energy flux at each scales are displayed in Fig. \ref{fig:flux} (Run \nbRoman{1}). 
It shows the relative proportion of energy flowing through the velocity and magnetic non-linear channels. Despite the fact that the same amount of kinetic and magnetic 
energy is injected in the system, on can see that the velocity flux 
$\Pi_{uu}^{u}$ channel carries $\sim 3$ times more energy than the Hall $-d_{i}\Pi_{bj}^{b}$ one.  Consequently, the dissipation rate of kinetic energy
$(\Pi^{u}_{diss}(\textbf{k})=   2\nu_{3}\sum_{\textbf{k}^{'}=0}^{\textbf{k}^{'}=\textbf{k}}\textbf{k}^{'6}E^{u}(\textbf{k}^{'}))$ is about $3$ times higher than the dissipation 
rate of magnetic energy $(\Pi^{b}_{diss}(\textbf{k})=   2\eta_{3}\sum_{\textbf{k}^{'}=0}^{\textbf{k}^{'}=\textbf{k}}\textbf{k}^{'6}E^{b}(\textbf{k}^{'}))$. Interestingly in 
absence of a mean magnetic field, kinetic and magnetic energy fluxes are in equipartition (see \cite{Meyrand_Thesis}). 
Otherwise said the relative proportion of kinetic and magnetic energy transfer rates is directly proportional to 
the strength of the mean magnetic field. This suggests that the energy tends to follow the non-linear path that pits it the least resistance (i.e. for which the non-linear 
time scale is the smallest), which is somewhat the way an electric current is distributed in a circuit with different resistivity.  
This property may have a profound impact on the various kinetic heating processes in the solar wind. Note that this remarkable property explains also why it is 
difficult to reach a pure weak turbulence regime in Hall MHD (see the discussion in section \ref{section:domain_of_validity}). 

\begin{figure*}
       \subfigure[]{ \label{fig:transfers_u}\includegraphics[width=0.666\textwidth]{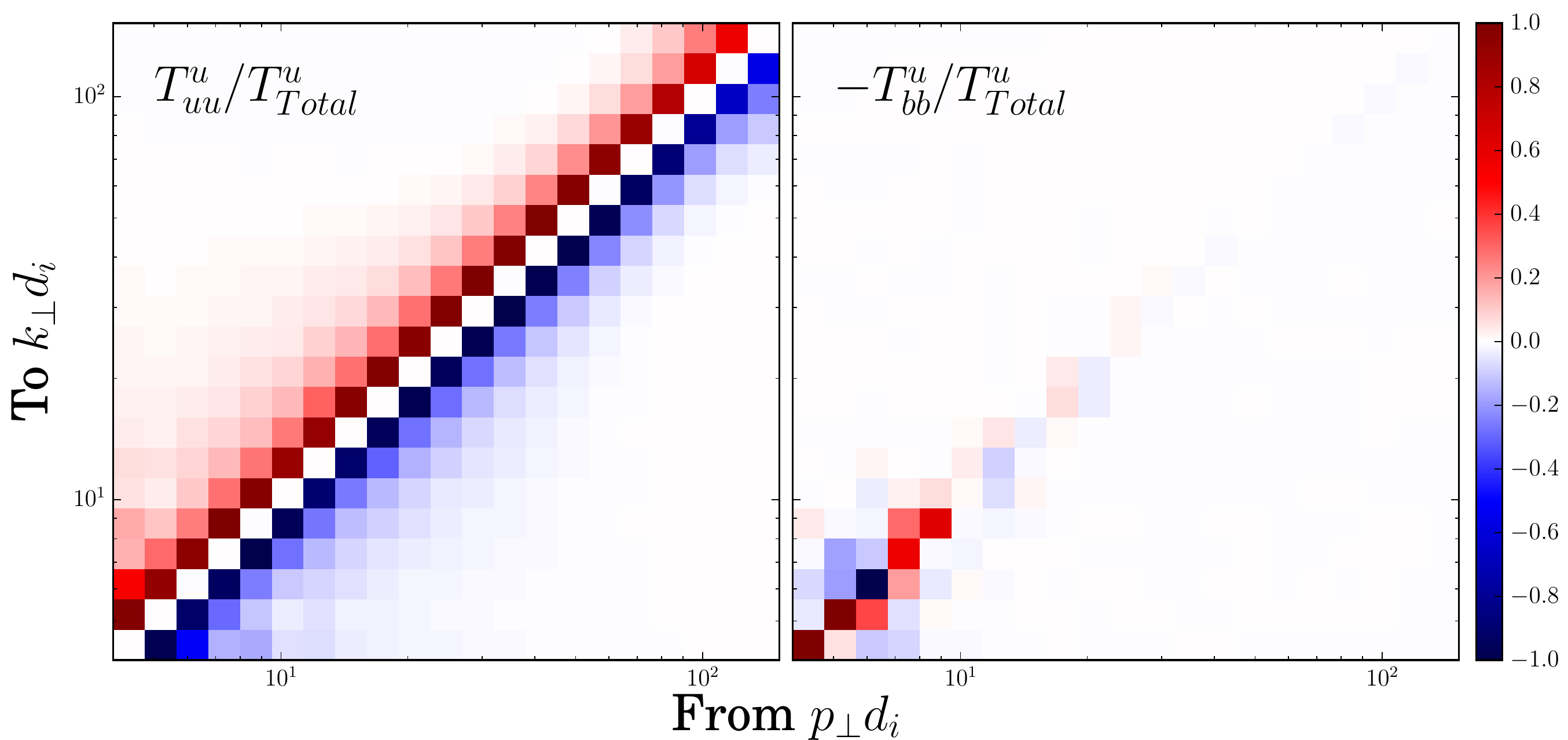}}   
        \subfigure[]{ \label{fig:transfers_b}\includegraphics[width=\textwidth]{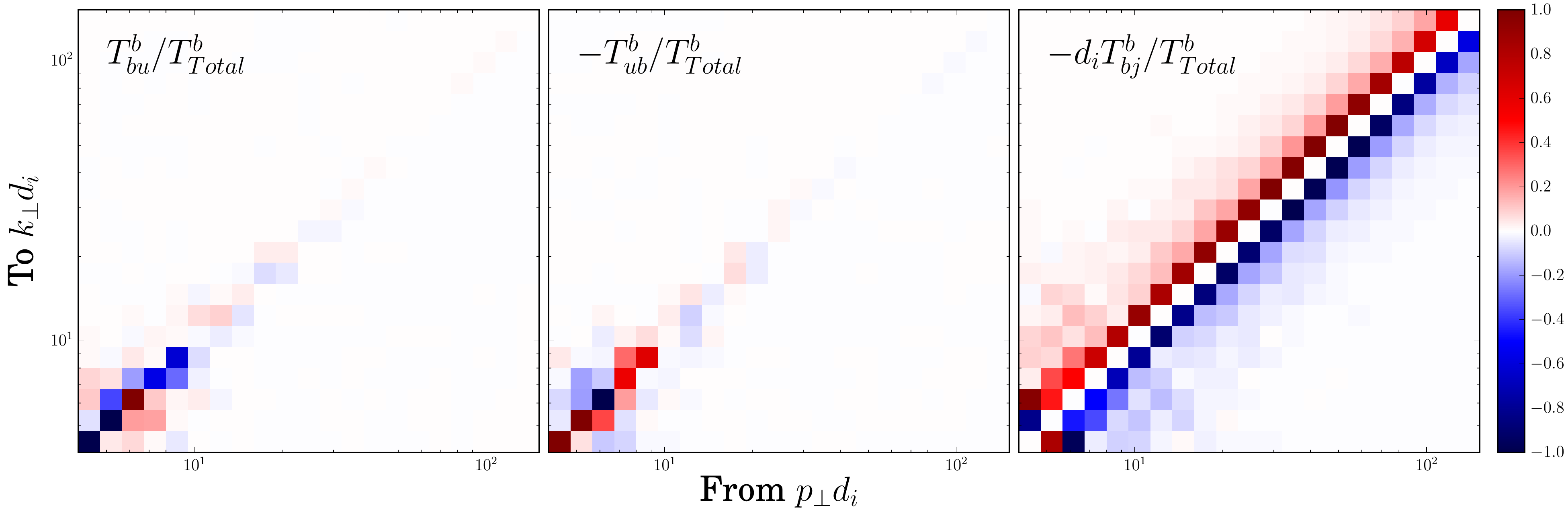}}
         \caption{Normalized transfer functions for the kinetic energy (top) and magnetic energy (bottom). Run \nbRoman{1}.}
         \label{fig:transfers}
\end{figure*}

The kinetic and magnetic energy transfer functions defined in equations (\ref{dtEu}), (\ref{dtEb}) and (\ref{transfers}) normalized respectively by the total kinetic and 
magnetic energy transfer functions are displayed in Fig. \ref{fig:transfers}. 
Clearly the Hall term is the dominant non-linear channel to cascade the magnetic energy toward small scales.
The kinetic energy on the other hand cascades predominately {\it via} the advection term $(\textbf{u}\cdot\nabla)\textbf{u}$. The latter property, given that the 
L-fluctuations do not belong to the weak turbulent regime for all $k_{\perp}$, may explain the $k_{\perp}^{-5/3}$ spectrum observed for the L-fluctuations velocity 
(see Fig. \ref{fig:spectrum_u}). 
The signals relative to $-d_{i}T_{bj}^{b}$ and $T_{uu}^{u}$ being concentrated around the diagonal $k_{\perp} = p_{\perp}$ means that direct and local energy transfers 
dominate. We can therefore confirm that the non-locality of the energy cascade is not responsible for the anomalous anisotropy observed in our simulations (see 
discussion in section \ref{section:spectra}).  
The cross transfers of energy from kinetic to magnetic fields $T_{bu}^{b}$ as well as from magnetic to kinetic fields $T_{bb}^{u}$ mediated respectively by the non-linear 
term $(\textbf{b}\cdot\nabla)\textbf{u}$ and $(\textbf{b}\cdot\nabla)\textbf{b}$ become negligible for scales $k_{\perp}d_{i}> 10$. Remarkably, this critical scale 
corresponds precisely to the one for which we observed a knee in the R-magnetic energy spectra and is close to the critical scale corresponding to the transition from weak 
ion cyclotron wave turbulence to strong ion cyclotron wave turbulence. This suggests that the the anomalous spectrum in $k_{\perp}^{-2.8}$ 
observed at $k_{\perp}d_{i}<10$ is due to the influence of the ion-cyclotron dynamics whereas the subsequent $k_{\perp}^{-2.5}$ spectrum may correspond to a pure 
weak whistler wave turbulence regime. 
Although this conclusion may seem appealing, it remains to show that resonant whistler three-wave interaction processes, the 
\q{atom} of the statistical WTT, are effectively at work. This is absolutely not guaranteed given that the whistler waves are embedded in a sea of highly non-linear 
ion-cyclotron fluctuations. The purpose of the following section \ref{sec5} is precisely to address this delicate issue.

\section{Measurements and characterisation of the resonant non-linear wave interactions}\label{sec5}
Even though the Hall MHD fluid equations are deterministic, turbulence is fundamentally a chaotic motion. Consequently, only averaged quantities are  experimentally 
reproducible and can be studied thoroughly. However, by adopting a statistical description we are paying an heavy price as we have to handle an infinite system of 
equations. In WTT the infinite hierarchy of equations is closed because in the long-time limit, the non-linear regeneration of third order moments depends essentially on 
products of second order moments and not on the fourth order cumulant which is not a resonant term \cite{Galtier09}. Consequently, the non-linear energy transfer in WTT  
involves mainly resonant three-wave interaction processes. 
Thenceforth, it appears essential to check whether such a process is effectively at work in our simulations before using 
the theoritical WTT framework. To do so it is necessary to use higher-order polyspectra which can be seen as a generalization of the Fourier analysis to include 
information about phase coherence.
\subsection{Definition of bispectra and physical interpretation}
A linear system can be described by a superposition of statistically independent Fourier modes and all the relevant information is contained in the power spectral density 
(or the autocorrelation function). However, if there exists some non-linear physical processes, then the phases of the Fourier modes are not independent anymore and 
information is also conveyed by the phases. By construction second-order statistics are phase blind and information about phase must be recovered from higher-order 
polyspectra. The use of higher-order moments nullifies all Gaussian random effects of the process, 
and the bispectrum can then quantify the degree of the remaining non-linear coupling. The bispectrum is defined by
 \begin{equation}
B(\omega_1, \omega_2) = \langle a(\omega_1)a(\omega_2)a(\omega_1 +\omega_2)^{*}\rangle
\end{equation}
where the average $\langle ... \rangle$ stands for averaging over time windows and $a$ represents the Fourier transform in time of the physical quantity of interest. The 
bispectrum measures the amount of phase coherence between three Fourier modes that obey the frequency summation rule $\omega_{1}+\omega_{2} = \omega_{3}$; 
it can be seen as the frequency domain representation of the third-order cumulant, the building block of WWT. The bicoherence 
\begin{equation}
C^{2}(\omega_{1}, \omega_{2}) = \dfrac{\vert\langle a(\omega_1)a(\omega_2)a(\omega_1 +\omega_2)^{*} \rangle\vert^{2}}{\langle \vert a(\omega_1)a(\omega_2) 
\vert^2 \rangle\langle \vert a(\omega_1 +\omega_2)^{*} \vert^2 \rangle}
\label{bicoherence}
\end{equation}
removes the magnitude dependence of the bispectrum, it is its normalized representation. With such a normalization, the coherence lies between 0 (no correlation) and 1 
(perfect correlation). If the three Fourier components $a(\omega_1)$, $a(\omega_2)$ and $a(\omega_1 +\omega_2)^{*}$ are phase locked they will sum without 
cancelling resulting in a large value of the bicoherence even though each phase, when taken separately, may vary in a random way.  The bicoherence measures therefore 
the proportion of the signal energy at any bifrequency  $(\omega_1, \omega_2)$ that is quadratically phase coupled to $\omega_{3}= \omega_1 + \omega_2$. Note that 
it is necessary to have a large time separation between the phases and the amplitudes characteristic times to produce a significant bicoherence. The use of bicoherence is 
therefore particularly relevant for wave phenomena with a weak departure from non-linearity and is a natural statistical tool to test the adequacy of weak interaction 
theory as a description of the non-linear coupling.  
Bicoherence have been widely used to examine various physical systems including space plasma physics \cite{Bale96, 
Dudok95}, plasma fusion device \cite{Ritz88, Diamond00}, ocean waves \cite{Mccomas80}, weak turbulence of gravity-capillary waves \cite{Aubourg15} or cosmology 
\cite{Verde02} to mention just but a few. For the sake of clarity we have defined bicoherence by considering a single quantity $a$ but bicoherence can easily be extended 
to study the phase coupling between different variables. To distinguish the two situations one should use the prefix auto- for a single quantity and cross- for multiple 
quantities.
\begin{figure}
        \includegraphics[width=1.0\linewidth]{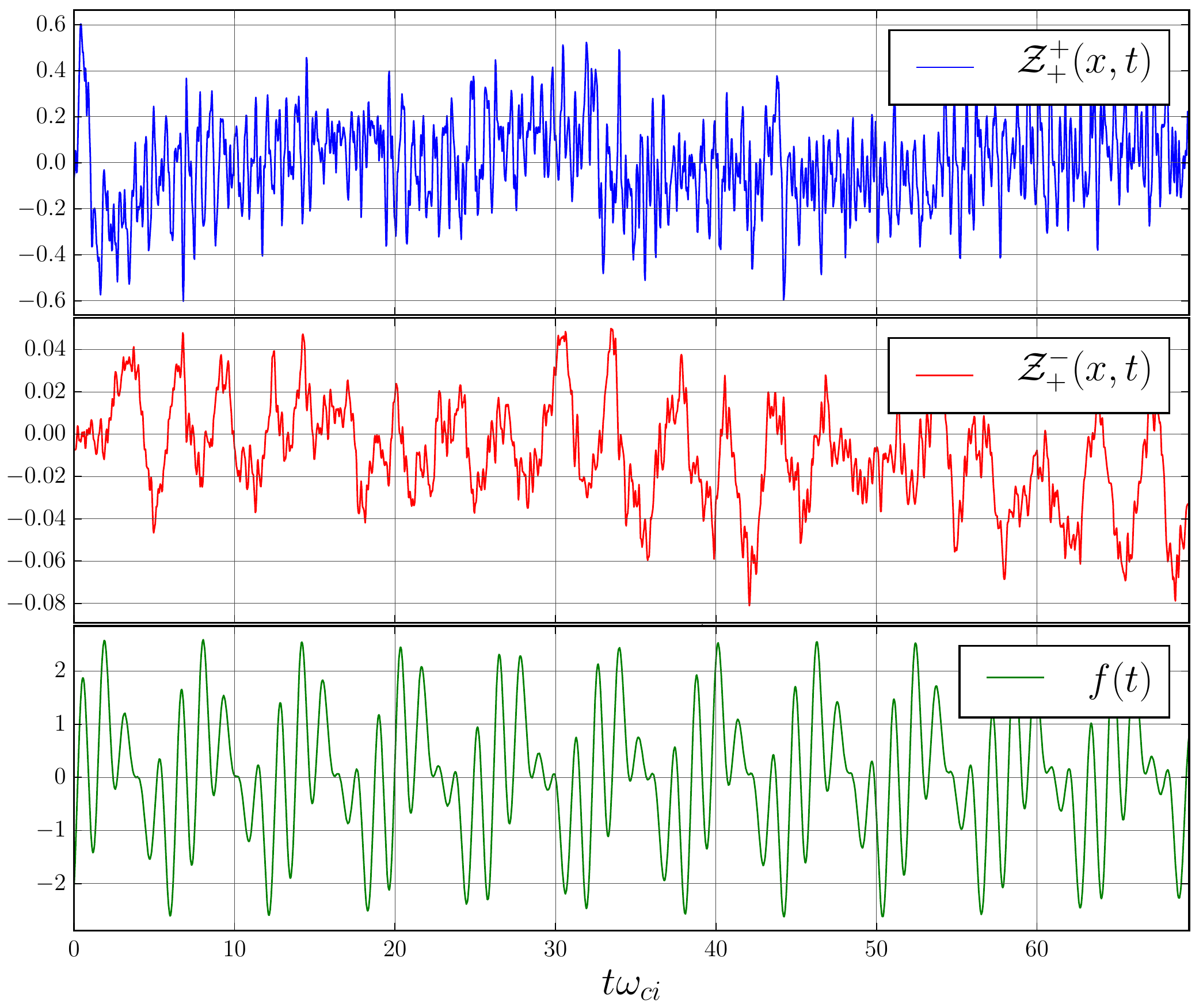}
    \caption{Temporal evolution of right ${\cal Z}_{+}^+(x,t)$ (top) and left ${\cal Z}_{+}^-(x,t)$ (middle) handed fluctuations as well as the synthetic fiducial signal 
    (bottom; see text) over one of the 100 windows (Run \nbRoman{3}).}
    \label{fig:time_serie}
\end{figure}

\subsection{Experimental setup}
\label{Experimental setup}
The bicoherence is related to the shape (in a statistical sense) of the time series. For a finite time serie even a truly Gaussian process will have a non-zero 
bicoherence. To decrease the noise level under the physically pertinent signal it is therefore necessary to consider a large statistical ensemble and thus to integrate the 
Hall MHD equations over a long time period. This turns out to be prohibitory for the first numerical experiment (Run \nbRoman{1}). We therefore consider a numerical simulation 
with a smaller resolution (Run \nbRoman{3}) of  $N_{\perp}^{2}\times N_{\parallel} = 128^2\times 64$ collocation points. The hyperdiffusivity and hyperviscosity are adjusted 
consequently with values of respectively $\eta_{3} = 3.2 \times 10^{-8}$ and $\nu_{3}=4.25 \times 10^{-9}$. All other parameters are otherwise identical to those of  
Run \nbRoman{1}. We have checked (not shown) that this numerical experiment presents qualitatively a similar \q{anomaly} at the level of the anisotropy 
and the power spectral index. However, because of the reduced spectral resolution, the wavenumber extension where highly non-linear left handed fluctuation are 
observed is reduced compare to Run \nbRoman{1}.

We record a time interval of the generalized Els\"asser fields ${\cal Z}_{\Lambda}^s(x,t)$ from a real space Eulerian probe. The modes $\vert k \vert \in [0, 
4d_{i}]$ are filtered so as to avoid any integral scale effect.  We consider a time interval of $6900\omega_{ci}$ with a sampling frequency of $\delta t\sim 
0.014\omega_{ci}$. From ${\cal Z}_{\Lambda}^s(x,t)$, we compute the Fourier transform in time over 100 time windows to obtain ${\cal Z}_{\Lambda}^s(x,\omega)$.  
Before taking the Fourier transform each sample is multiplied by a Hamming window and detrended using a standard linear least-squares method. Cross-bicoherence  are 
then computed from equation (\ref{bicoherence}). 
Furthermore we calculate a fiducial bicoherence from 100 synthetic signals $f(t) = \sum_{i=1}^{3}\text{sin}\left( \left( \omega_{i}+\delta \omega_{i}\right) 
t+\phi_{i}\right)+\delta a t$ with similar sampling frequency and where $\omega_{1} = 10.4 \omega_{ci}$,  $\omega_{2} = 2.5 \omega_{ci}$, and $\omega_{3} = 
\omega_{1} + \omega_{2}$.  The phases $\phi_{i}$ are randomly distributed over each time window such that $\phi_{i} \in \left[-\pi, \pi \right]$ and $\phi_{3} = \phi_{1} 
+ \phi_{2}$. $\delta \omega_{i}$ are random numbers modelling artificial non-linear frequency broadening with $\delta \omega_{i} \in \left[-0.001, 0.001 \right]$. Finally, 
$\delta a$ simulates a Gaussian noise at the level of the amplitude ($\delta a \in \left[-0.01, 0.01 \right]$). This fiducial bicoherence gives an idea of the statistical noise 
inherent to the statistical ensemble that we consider as well as the signature of quadratically phase coupled waves. 

\subsection{Results}
Figure \ref{fig:time_serie} displays the temporal evolution of right ${\cal Z}_{+}^+(x,t)$ and left ${\cal Z}_{+}^-(x,t)$ handed fluctuations as well as the synthetic fiducial 
signal over one of the 100 windows. Clearly, the R fluctuations evolve on shorter time scale than the L fluctuations. Interestingly the later displays a periodicity of $\sim 3 
\omega_{ci}^{-1}$ but shorter characteristic times are also excited. The use of bicoherence will show whether these latest short time fluctuations result from the coupling 
with whistler waves or are the consequence of non-linear processes involving exclusively ion cyclotron type fluctuations. 

\begin{figure*}
 \centering
    \subfigure[]{\label{fig:pp_mm_mm}\includegraphics[width=0.49\textwidth]{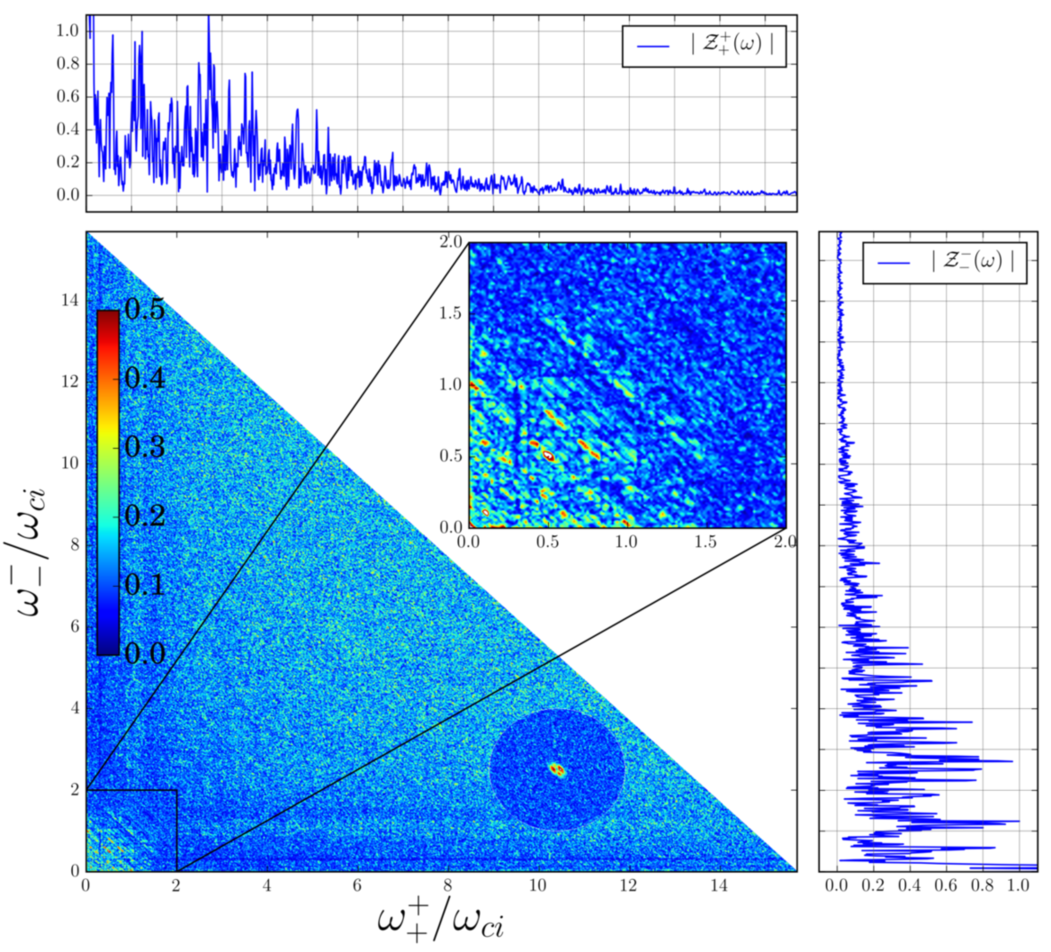}}
    \subfigure[]{\label{fig:pm_mp_pm}\includegraphics[width=0.49\textwidth]{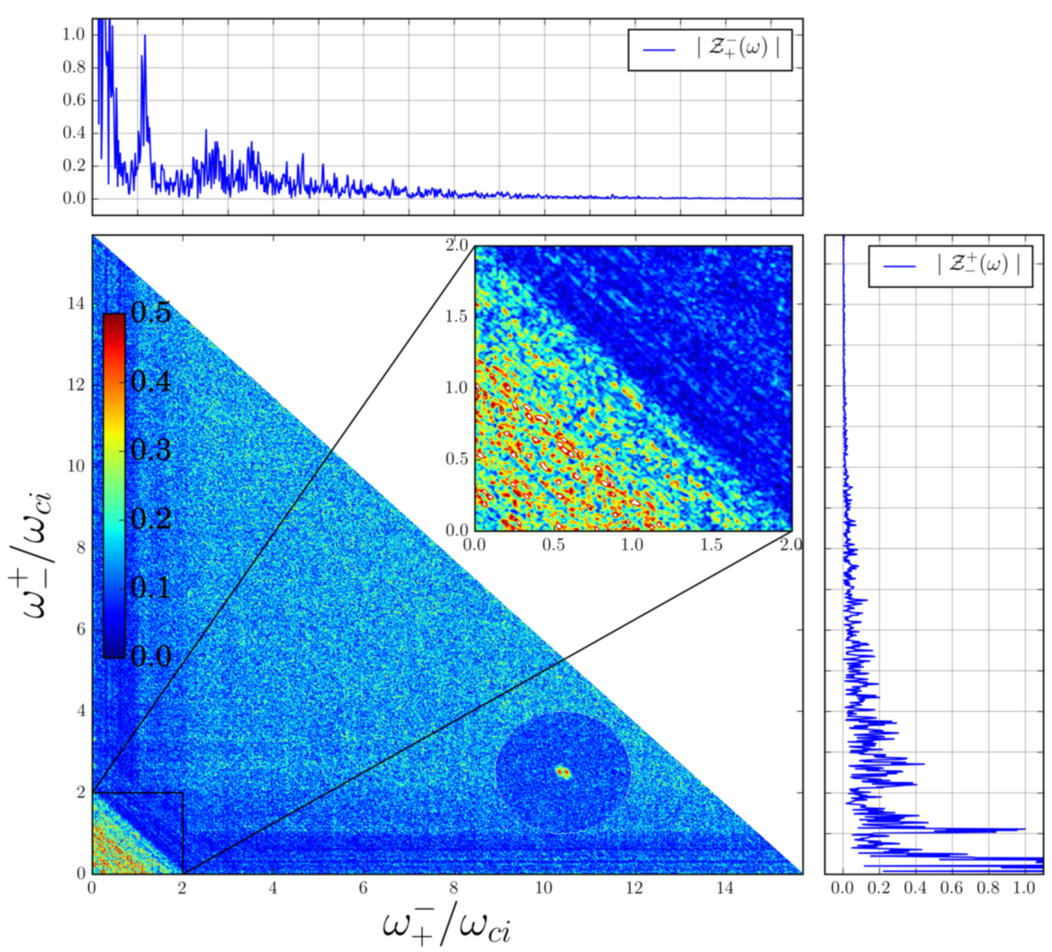}}
    \caption{\subref{fig:pp_mm_mm}: Cross-bicoherence $C_{RR}$ resulting from the coupling between three whistler waves whose two have opposite polarity.
\subref{fig:pm_mp_pm}: Cross-bicoherence $C_{LL}$ resulting from the coupling between three ion cyclotron waves whose two have opposite polarity.
For convenience the modulus of the time Fourier transform of the generalized  Els\"asser fields normalized by their respective maximum value around $\omega_{ci} = 1$ 
are also shown above and beside each plot.}
    \label{bicoh}
\end{figure*}  
Figure \ref{fig:pp_mm_mm} displays the cross-bicoherence 
\begin{equation}
C^{2}_{RR}(\omega_{1}, \omega_{2}) = \dfrac{\vert\langle {\cal Z}_{+}^+(\omega_1){\cal Z}_{-}^-(\omega_2){\cal Z}_{+}^+(\omega_1 +\omega_2)^{*} \rangle\vert^{2}}
{\langle \vert {\cal Z}_{+}^+(\omega_1){\cal Z}_{-}^-(\omega_2) \vert^2 \rangle\langle \vert {\cal Z}_{-}^-(\omega_1 +\omega_2)^{*} \vert^2 \rangle}
\label{bicoh1}
\end{equation}
resulting from the coupling between three whistler waves whose two have opposite polarity. Figure \ref{fig:pm_mp_pm} displays a similar cross-bicoherence but for ion 
cylotron waves,  
\begin{equation}
C^{2}_{LL}(\omega_{1}, \omega_{2}) = \dfrac{\vert\langle {\cal Z}_{+}^-(\omega_1){\cal Z}_{-}^+(\omega_2){\cal Z}_{+}^-(\omega_1 +\omega_2)^{*} \rangle\vert^{2}}
{\langle \vert {\cal Z}_{+}^-(\omega_1){\cal Z}_{-}^+(\omega_2) \vert^2 \rangle\langle \vert {\cal Z}_{+}^-(\omega_1 +\omega_2)^{*} \vert^2 \rangle}.
\label{bicoh2}
\end{equation}
For convenience, the modulus of the time Fourier transform of the generalized Els\"asser fields relative to one time window normalized by their respective maximum value 
around $\omega_{ci} = 1$ are also shown.  The Nyquist theorem restricts the displays of the cross-bicoherence to the triangle defined by $\omega_{1}+\omega_{2} \leq 
\text{nf}$, where $\text{nf}\sim 15 \omega_{ci}$ is the Nyquist frequency. The circular domain located at the bottom left of the figures corresponds to the fiducial 
bicoherence described in section \ref{Experimental setup}. 

Significant cross-bicoherence magnitude emerges from the statistical convergence noise at all frequencies which confirms that three-wave resonant processes are indeed 
present in the signal. A noticeable organisation of the cross-bicoherence is clearly visible. The cross-bicoherence level is not homogeneous showing preferential 
interaction among the waves. Not surprisingly the highest value for ion cyclotron cross-bicoherence $C_{LL}$ is seen for values of $\omega_{1}$, $\omega_{2}$ such that  
$\omega_{1}+\omega_{2} \leq 2\omega_{ci}$ (see inset on Fig. \ref{fig:pm_mp_pm}).  The fact that significant cross-bicoherence is found for $\omega_{1}, \omega_{2} 
\in \left[\omega_{ci}, 2\omega_{ci}\right]$ 
may be attributed to non-linear frequency broadening.  Interestingly this observation is also valid for whistler cross-bicoherence $C_{RR}$ and suggests that whistlers 
are generated by non-linear interactions involving ion cyclotron waves. 
Conversely, it can be observed in Fig. \ref{fig:pm_mp_pm} that the ion cyclotron cross-bicoherence $C_{LL}$ above the statistical noise is present at frequency $\omega > 
\omega_{ci}$ far from the linear frequency asymptote. This suggests that ion cyclotron fluctuations may be generated by non-linear interactions involving whistlers. To test 
this idea we have computed the cross-bicoherence $C_{LR}$ resulting from the coupling between an ion cyclotron wave and two counter-propagating whistler waves with
\begin{equation}
C^{2}_{LR}(\omega_{1}, \omega_{2}) = \dfrac{\vert\langle {\cal Z}_{-}^+(\omega_1){\cal Z}_{+}^+(\omega_2){\cal Z}_{-}^-(\omega_1 +\omega_2)^{*} \rangle\vert^{2}}
{\langle \vert {\cal Z}_{-}^+(\omega_1){\cal Z}_{+}^+(\omega_2) \vert^2 \rangle\langle \vert {\cal Z}_{-}^-(\omega_1 +\omega_2)^{*} \vert^2 \rangle} .
\label{bicoh3}
\end{equation}

\begin{figure}
       \includegraphics[width=1.0\linewidth]{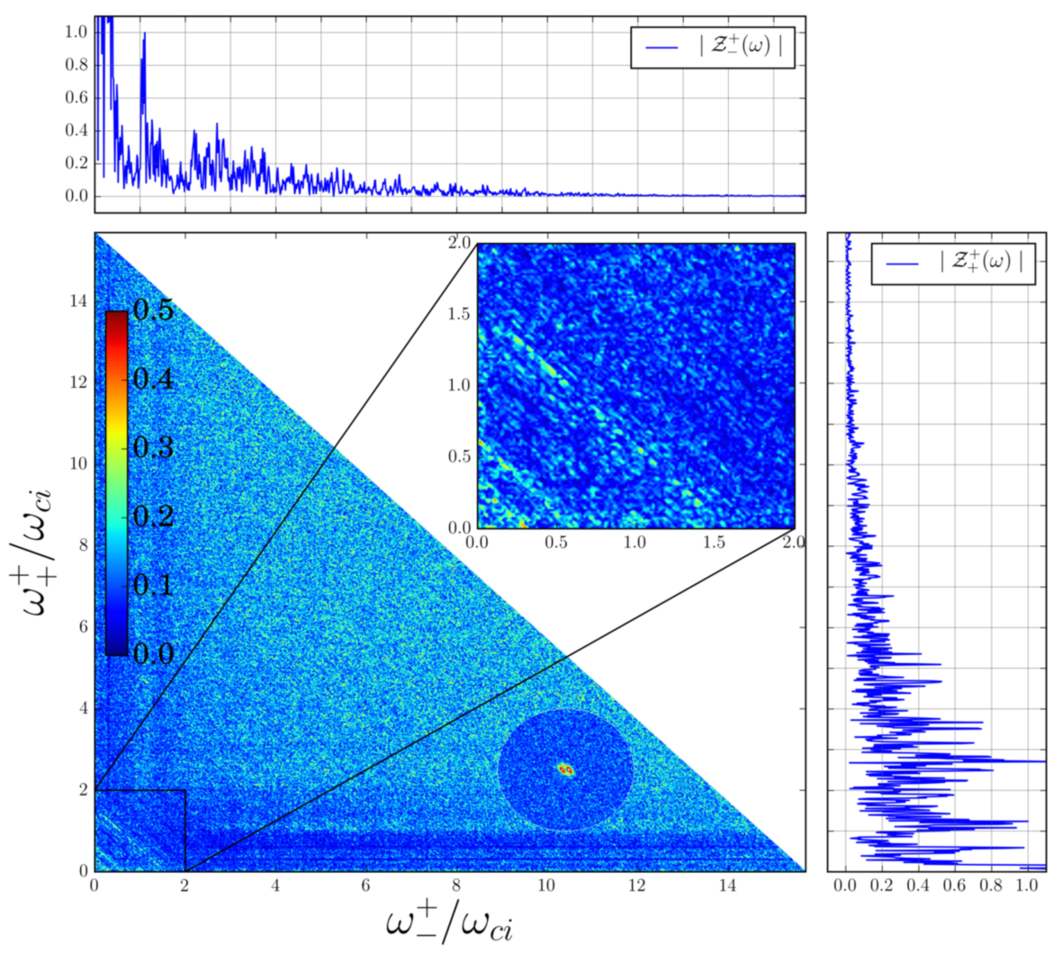}
        \caption{Cross-bicoherence $C_{LR}$ resulting from the coupling between an ion cyclotron wave and two counter-propagating whistler waves. For convenience the 
        modulus of the time Fourier transform of the generalized  Els\"asser fields normalized by their respective maximum value around $\omega_{ci} = 1$ are also 
        shown above and beside the plot. }  
         \label{fig:mp_pp_mm} 
\end{figure}
Figure \ref{fig:mp_pp_mm} displays such a cross-bicoherence and confirms the result obtained from the detailed study of the transfer functions provided in section 
\ref{section:transfers}. Ion cyclotron and whistler waves are indeed non-linearly entangled. Note that all the different combinations of cross-bicoherence are qualitatively
similar (not shown). However, we observed that the cross-bicoherence involving at least two counter-propagating waves is larger than the corresponding cross-bicoherence 
for which only waves propagating in the same direction take part. 

The new and important information that we can extract from the cross-bicoherence analysis is that the resonant triadic interactions survive the highly non-linear bath of ion 
cyclotron fluctuations. However, this coupling is significantly higher in the frequency domain where only weakly non-linear waves coexist namely for $\omega_{1}$,
$\omega_{2}$ such that  $\omega_{1}+\omega_{2} \leq 2\omega_{ci}$. This observation suggests an alteration of the resonant triadic interactions by the presence of the 
strong wave ion cyclotron turbulence.
In order to study deeper the non-linear dynamics we focus in the following section on the properties of the space-time Fourier spectrum. Such a technique allows the 
precise identification and extraction of the waves and non-linear structures contributions to the total energy. 
\section{Space-time Fourier spectra}\label{sec6}
\subsection{Experimental setup}
Computation of the space-time Fourier  ($\textbf{k}-\omega$) spectrum requires simultaneous space and time Fourier transforms. The frequency sampling must be at 
least two times larger than the frequency of the fastest waves of the system, and the total time of acquisition should be larger than both the period of the slowest waves 
and the turnover time of the slowest eddies. These constraints turn out to be numerically redhibitory for Run \nbRoman{1}. We thus consider a numerical simulation 
with a smaller resolution (Run \nbRoman{4}) of  $N_{\perp}^{2}\times N_{\parallel} = 256^2\times 128$ collocation points. The hyperdiffusivity and hyperviscosity are 
adjusted consequently with values of respectively $\eta_{3} = 4.5 \times 10^{-10} $ and $\nu_{3}=7.6 \times 10^{-11}$. All other parameters are otherwise identical to those 
of Run \nbRoman{1}. We have checked (not shown) that this numerical experiment presents qualitatively a similar \q{anomaly} at the level of the 
anisotropy and the power spectral index. However, because of the reduced spectral resolution, the wavenumber extension where highly non-linear left handed fluctuations 
are observed is reduced compare to Run \nbRoman{1}.
Because the dispersion relation of incompressible Hall MHD depends on the angle $\theta = \text{arccos}(k_{z}/\textbf{k})$ of the wavenumber relative to the mean field 
(see section \ref{Eigenmodes}), we consider the Fourier transform of the generalized  Els\"asser fields along rays of wavenumbers at constant $\theta$. We consider five 
different angles $\theta_{i}$ such as their cosines are equal to $0.1,0.3,0.5,0.7,0.9$. 
For each of these five angles we have considered 10 rays at constant $\varphi$ where $\varphi=\text{arctan}(k_{y}/\textbf{k})$ is the angle between the wavevector and the 
direction $y$ perpendicular to the mean field. We took $\varphi_{i} =-\pi/2 + i\times\pi/10$ for $i\in\mathbb{N} ~\vert~ i\in [0,9]$. Negative values of the wavenumber 
$k_x$ are not considered because of the symmetry of the Fourier representation of real fields ($u(-k)=u^{*}(k)$). 
Along each of these $10\times 5$ wavevector rays, we recorded time evolution of the generalized  Els\"asser fields on $64$ points uniformly distributed ($k_{j} = j\times 
d_{i} ~\vert~ j \in\mathbb{N} ~\vert~ j\in [1,64]$). Because these wavenumbers do not match the grid points, such procedure requires 3D Fourier space interpolation. We 
have opted for a cubic spline technique.  Due to simulation storage constraints (both in space as well as in I/O speed) we record only the real part of the x component of 
the generalized  Els\"asser field. We have checked that another choice does not change qualitatively the result. 
We choose an acquisition frequency equal to $dt = 4\times 10^{-3}\omega_{ci}$ which allows to correctly resolve the fastest waves that we observe in the 
simulation. The total time of acquisition is equal to $t \sim 60 \omega_{ci}$ which enables to capture both the ion cyclotron waves as well as the slowest eddies. In order to 
mitigate spectral leakage we applied systematically a Hamming window and substracted the mean value of the different signals before computing the temporal Fourier 
transform.
\subsection{Results}
\begin{figure}
       \includegraphics[width=1.0\linewidth]{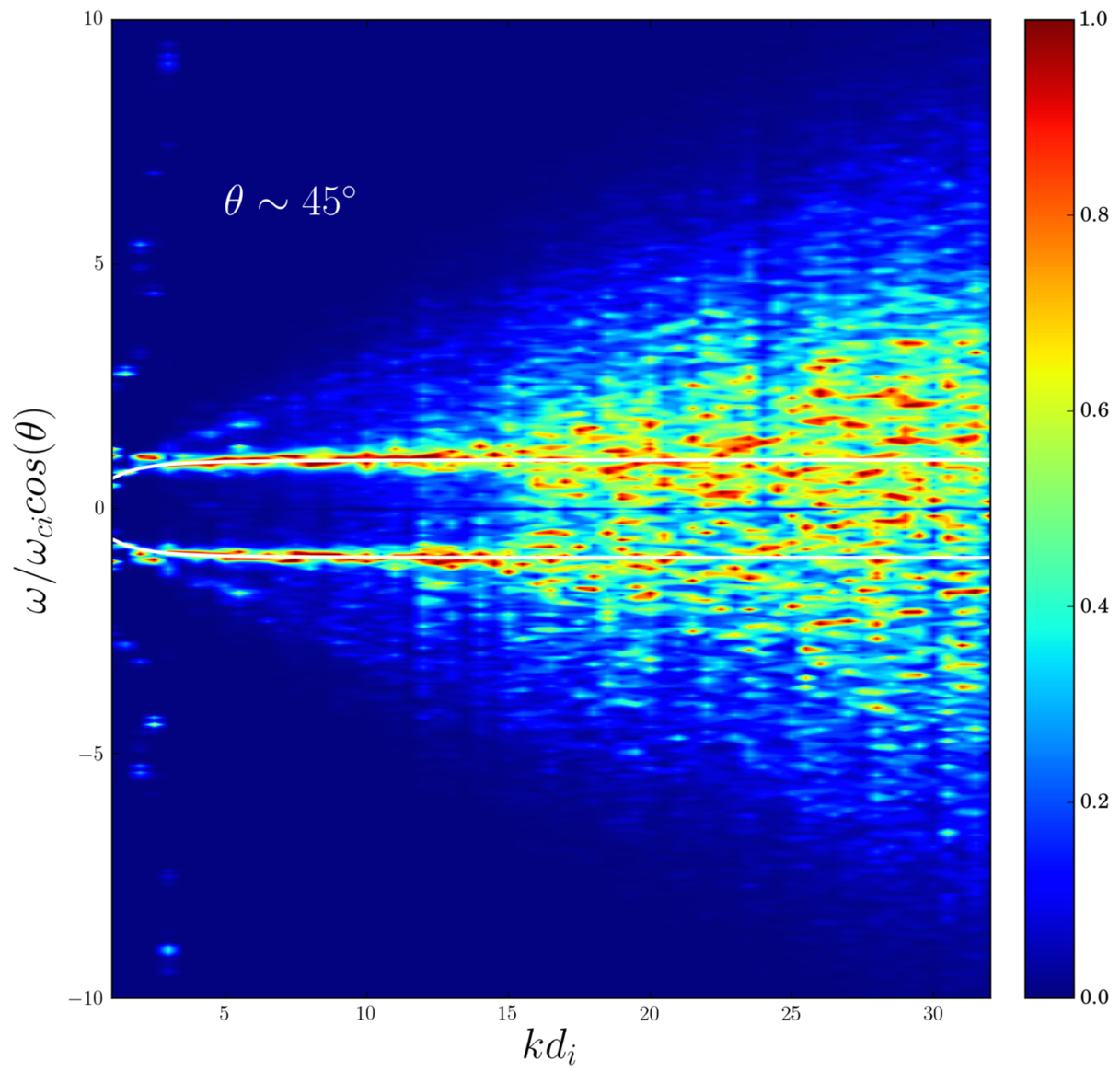}
        \caption{Space-time Fourier spectrum of the $x$ component of the generalized  Els\"asser $Z_{-}^{+}$ (positive frequency) and  $Z_{+}^{-}$ (negative 
        frequency) fields at $\theta \sim 45^{\circ}$. The solid white line is the theoretical ion cyclotron linear dispersion relation. The color map is normalized to the 
        maximum value of the spectrum at each fixed $k$. }  
    \label{fig:kw_ic}  
\end{figure}
\begin{figure*}
 \centering
    \subfigure[]{\label{fig:kw_w}\includegraphics[width=0.49\textwidth]{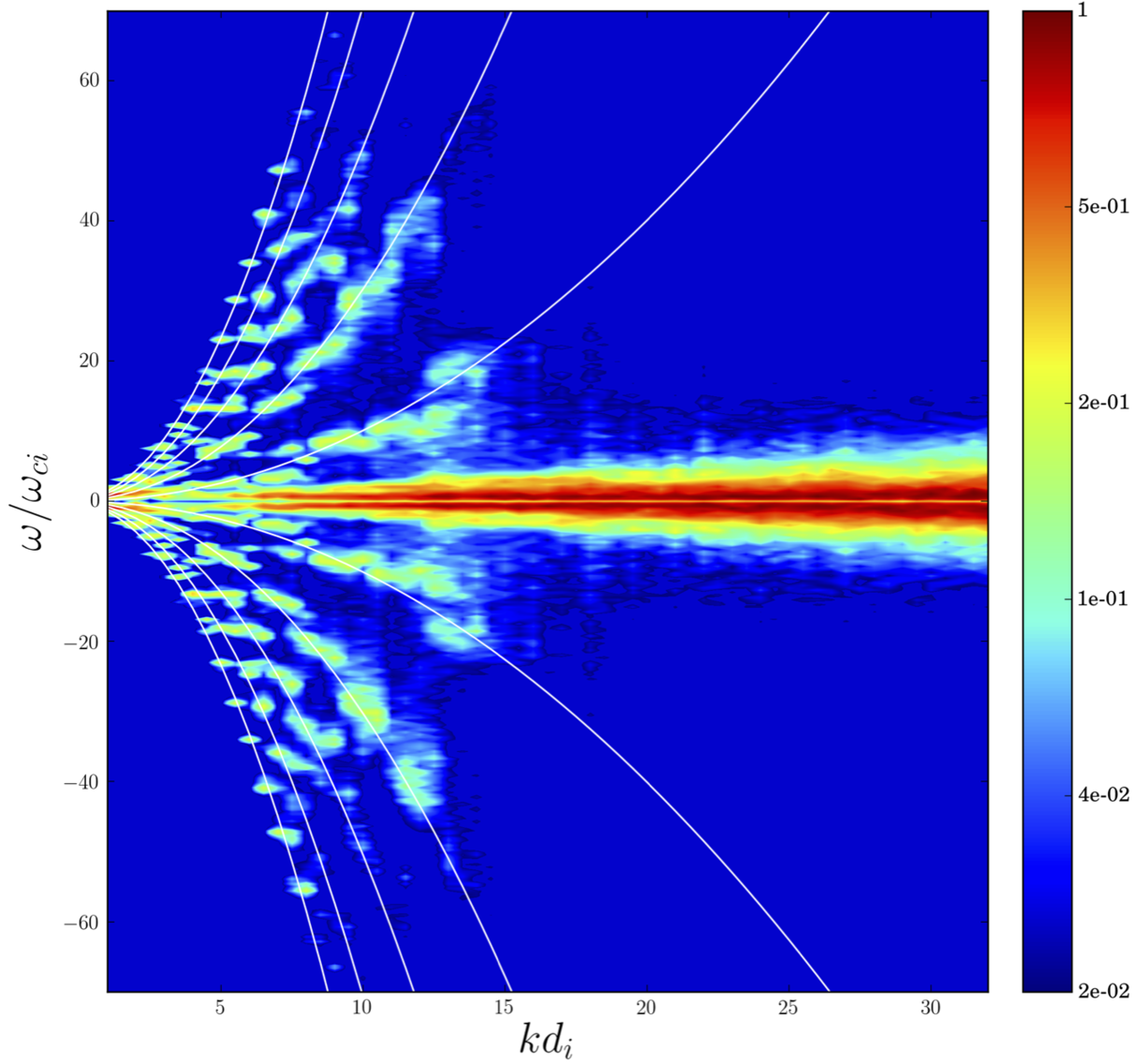}}
    \subfigure[]{\label{fig:kw_emhd}\includegraphics[width=0.49\textwidth]{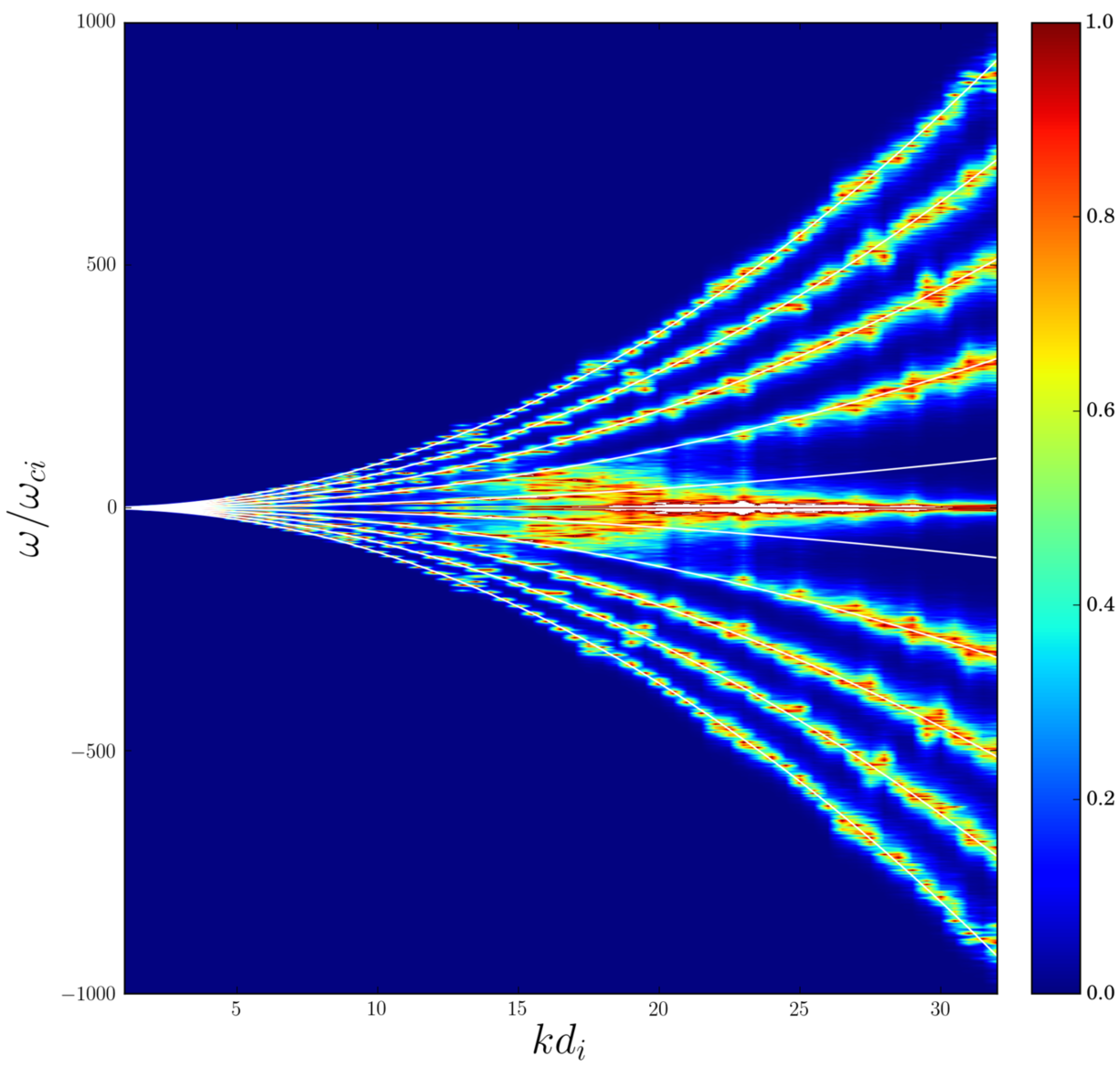}}
    \caption{\subref{fig:kw_w}: Mean of five space-time Fourier spectra of the $x$ component of the generalized  Els\"asser variables $Z_{+}^{+}$ (positive frequency) and  $Z_{-}^{-}$ (negative frequency) corresponding to five different angles $\theta \sim 84^{\circ}, 72^{\circ}, 60^{\circ}, 45^{\circ}, 25^{\circ}$ (run \nbRoman{4}). Before the mean is taken, each signal is normalized to the maximum value at each fixed $k$. The color scale is $\text{log}_{10} Z_{\Lambda}^{s}(k,\omega)$. Solid white lines are theoretical whistler linear dispersion relations corresponding to the five angles. 
\subref{fig:kw_emhd}: Same as \ref{fig:kw_w} but for the EMHD run \nbRoman{5} ($\textbf{u}=0$). Note that contrary to figure \ref{fig:kw_w} the color scale is linear,  frequency-scales are also different.}
\end{figure*}   
We show in Fig. \ref{fig:kw_ic} the space-time Fourier spectrum of the $x$ component of the generalized  Els\"asser variables $Z_{-}^{+}$ and $Z_{+}^{-}$ (ion cyclotron) 
corresponding to $\cos(\theta)=0.7$. At large scale ($kd_{i} < 15$), energy is mainly localized on the dispersion relation.  At smaller scales, one can observe a significant 
broadening of the $(\omega-kd_{i})$ distribution reflecting the transition toward strong ion cyclotron wave turbulence. The signal at high frequency ($\omega/ \omega_{ci} > 1$) and low wavenumber ($k<5$) matches the linear dispersion relation of the whistler and is a direct signature of the cross-coupling of the two populations of waves.
Figure \ref{fig:kw_w} displays the mean of five space-time Fourier spectra of the $x$ component of the generalized  Els\"asser variables $Z_{+}^{+}$ and $Z_{-}^{-}$ 
corresponding to five different angles $\theta \sim 84^{\circ}, 72^{\circ}, 60^{\circ}, 45^{\circ}, 25^{\circ}$. Before the mean is taken, each signal is normalized to the 
maximum value at each fixed $k$.  At large scale ($kd_{i} < 15$), energy is mainly localized on the dispersion relations. The weak non-linear effects manifest themselves in a broadening of the $(\omega-kd_{i})$ distribution with respect to the linear dispersion relations. Not surprisingly the broadening increases significantly as the angle 
approach $90^{\circ}$ for which the linear terms vanish. Interestingly, we also see around $\omega = \omega_{ci}$ significant energy which reflects the cross coupling 
with the ion-cyclotron waves. Beyond $kd_{i}\sim 15$, all the linear high frequency signal collapses toward low frequency. The fact that this phenomenon occurs at scales 
similar to those for which we observe the transition from weak to strong wave ion cyclotron turbulence suggests that the whistlers are locally (in $k$) killed by the strongly 
non-linear ion-cyclotron fluctuations. To test this idea, we have performed an EMHD simulation  ($\textbf{u}=0$) everything being equal otherwise (Run \nbRoman{5}). 
Figure \ref{fig:kw_emhd} is the EMHD version of Fig.\,\ref{fig:kw_w} corresponding to Run \nbRoman{5}. The $x$ component of the generalized  Els\"asser variables 
$Z_{+}^{+}$ and $Z_{-}^{-}$ are respectively replaced by the $x$ component of $\mathcal{B}_{+}$ and $\mathcal{B}_{-}$. The acquisition frequency is divided by 10 with 
respect to Run \nbRoman{4} ($dt = 4\times 10^{-4}\omega_{ci}$) in order to capture the fastest waves that we observe in this numerical experiment. The result differs 
strikingly compared to the Hall MHD simulation. We want to draw the reader's attention to the fact that the frequency-scales of Fig. \ref{fig:kw_w} and \ref{fig:kw_emhd} are different. The whistler fluctuations spread in all the linearly accessible $k-\omega$ space and demonstrates indirectly that the local (in $k$) non-linear cross-coupling 
between strongly non-linear ion cyclotron fluctuations and whistler waves tend to kill the latter. 
It interesting to note, however, that the modes corresponding to the angle $\sim 84^{\circ}$ do not follow the linear dispersion for all $k$. They experiment a transition 
toward strong turbulence and seem to affect slightly the modes corresponding to the angle $\sim 72^{\circ}$. They can be seen as a low frequency strongly non-linear 
quasi-2D condensate. Remarkably these strongly interacting wave modes, which are ineluctable as the weak turbulence dynamics develops a strong anisotropy 
\cite{Meyrand13}, do not affect significantly the dynamics of the weakly interacting one's. This situation has been also observed in weak MHD wave turbulence 
\cite{Meyrand15,Meyrand16} and is of primary importance as it validates \textit{a posteriori} the WTT approach for whistler turbulence.
\section{Conclusion} \label{sec7} 
\subsection{Summary}
It is now time to come back to the original motivation for the above developments. In this paper, we have considered magnetized plasma turbulence in the 
framework of incompressible Hall MHD by taking as a starting point the WTT \cite{Galtier06}. The confrontation of this theory with numerical experiments allowed 
us to highlight new and interesting properties. We have shown that the ion cyclotron and whistler waves populations are not transparent with respect to each other 
in contrast of what is implicitly assumed in the WTT when the power law spectrum solutions are derived. 

We have shown and explained why it is difficult if not impossible to reach a situation for which both populations of waves are weakly non-linear (Section 
\ref{section:domain_of_validity}). The standard situation in Hall MHD appears therefore to be a mixture of weakly interacting ion cyclotron and whistler waves 
embedded in a bath of higly non-linear ion cyclotron fluctuations. This situation has a profound impact on the statistical properties. It produces anomalous (with 
respect to the WTT) scaling and anisotropy (Section \ref{sec4}).  

Using higher-order polyspectra we have shown that resonant triadic interactions, the building block of WTT, survive the bath of highly non-linear ion cyclotron 
fluctuations (Section \ref{sec5}). This study allowed us to highlight the importance of resonant triadic interactions involving waves with different polarity. The 
solutions derived in Hall MHD WTT do not take into account this cross-coupling and need therefore to be amended consequently. 

The study of the space-time Fourier spectra enables us to look deeper in the non-linear dynamics (Section \ref{sec6}).  We have seen in particular that whistler waves 
and consequently the weak wave turbulent dynamics survive only in the $k$-space region where ion cyclotron waves are weakly interacting, a situation which 
limits significantly the domain of applicability of the WTT framework applied to Hall MHD. 
\subsection{Discussion}
When facing a question, which everyone agrees is extremely difficult to solve, it is always beneficial to reformulate it in simplified terms in order to begin to analyze it. In this 
spirit, incompressible Hall MHD is interesting to the extent that it is the simplest plasma model describing both large MHD scales and sub-proton scale dynamics of electrons. 
Nevertheless, it is imperative to come back to this simplification and analyse thoroughly its limitations. In particular, one must ask about the implications of our results might have 
in the context of astrophysical and space plasma turbulence, for which it is an observational certainty that turbulence develops at collisionless scales \cite{Bale05}. 

Weak collisionality implies that on the timescales of interest, a kinetic description that evolves the distribution functions of the particles is required. Fortunately, a strong 
magnetization induces a strong anisotropy which can benefit analytically to yield kinetic models that reduce the phase-space to only 5D \cite{Schekochihin09} or 
even 4D \cite{Zocco11}. Furthermore, when charged particles are magnetized, they do not need to exchange information about their mean perpendicular velocity by 
collisions to have a collective or fluid like behavior. They just have to \q{read} the local magnetic and electric fields. This basic property explains largely (and partly) 
the remarkable capacity of fluid models like reduced MHD \cite{Chen11} and reduced EMHD \cite{Boldyrev12} to account for phenomenon observed in collisionless 
plasmas. 

Using the Hall MHD model, we have implicitly assumed that ions have a fluid behavior at sub-proton scales even if, due to their large mass, they tend to be 
demagnetized at those scales. This is the main limitation of our approach if one wants to extrapolate our results to collisionless plasma turbulence. In fact, kinetic 
simulations that span the macroscopic fluid scales down to the motion of electrons show that the energy spectra of ion velocity field falls off abruptly at proton 
scales, not showing any clear power law at higher wavenumbers \cite{Karimabadi13,Franci15} in contrast to what we observed in our Hall MHD simulations. 
So, the question is: what are the lessons that can be drawn from our study in the context of collisionless plasma turbulence? 

The main message of the present work is that from the moment that different waves coexist in a turbulent environment, they are irredeemably 
interwoven by the non-linearities, even if they \q{live} in remote areas of the frequency space. As such, they lose their own identity since they exchange 
some of their characteristics. We believe that this situation have important implications and is to be taken into account if one wants to understand thoroughly 
magnetized-plasma turbulence. 
\section*{Acknowledgements} 
The research leading to these results has received funding from the European Commission's 7th Framework Program (FP7/2007-2013) under the grant agreement SHOCK 
(project number 284515), from the ANR contract 10-JCJC-0403 and ANR-JC project THESOW. The computing resources were made available through the UKMHD 
Consortium facilities funded by STFC grant number ST/H008810/1, the HPC resources of CCRT/CINES/IDRIS under allocation 2012 [x2012046736] made by GENCI and  the Savio computational cluster resource provided by the Berkeley Research Computing program at the University of California, Berkeley. R.M  acknowledges the financial support from EU-funded Marie Curie-Sk\l{}odowska Global 
Fellowship.
\bibliographystyle{apsrev4-1}
\bibliography{biblio}
\end{document}